\documentclass[article,amsmath,amssymb,superscriptaddress,12pt]{revtex4}

\usepackage{xcolor}
\usepackage{graphicx}
\usepackage{dcolumn}
\usepackage{bm}
\usepackage{fancybox}

\usepackage{color}

\usepackage{bm}%
\usepackage{graphicx}
\usepackage{amssymb,amsmath}
\usepackage{subfig}
\usepackage{physics}
\usepackage{cancel}

\usepackage[utf8]{inputenc}

\newcommand{\rr}{{\bm r}}

\renewcommand{\vec}{\bm}

\NewDocumentCommand{\orb}{ m m }{\phi_{#1}(\vec{#2})}

\usepackage{xparse}

\begin{document}


\title{Improving the exchange and correlation potential in density functional approximations through constraints}

\author{Timothy J.\ Callow}
\email{t.callow@hzdr.de}
\affiliation{Department of Physics, Durham University, South Road, Durham, DH1 3LE, United Kingdom}
\affiliation{Max-Planck-Institut f\"{u}r Mikrostrukturphysik, Weinberg 2, D-06120 Halle, Germany}

\author{Benjamin J.\ Pearce}
\email{b.j.pearce@durham.ac.uk}
\affiliation{Department of Physics, Durham University, South Road, Durham, DH1 3LE, United Kingdom}

\author{Tom Pitts}
\email{tom.pitts@durham.ac.uk}
\affiliation{Department of Physics, Durham University, South Road, Durham, DH1 3LE, United Kingdom}

\author{Nektarios N.\ Lathiotakis}
\email{lathiot@eie.gr}
\affiliation{Theoretical and Physical Chemistry Institute, National Hellenic Research Foundation, Vass. Constantinou 48, 116 35 Athens, Greece}

\author{Matthew J.\ P.\ Hodgson}
\email{matthew.j.hodgson@durham.ac.uk}
\affiliation{Department of Physics, Durham University, South Road, Durham, DH1 3LE, United Kingdom}

\author{Nikitas I.\ Gidopoulos}
\email{nikitas.gidopoulos@durham.ac.uk}
\affiliation{Department of Physics, Durham University, South Road, Durham, DH1 3LE, United Kingdom}

\date{\today}

\begin{abstract}
We review and expand on our work to impose constraints on the effective Kohn-Sham (KS) potential of local and semi-local density functional approximations. 
\color{black}
Constraining the minimisation of the approximate total energy density-functional invariably leads to an optimised effective potential (OEP) equation, the solution of 
which yields the KS potential.
We review briefly our previous work on this and demonstrate with numerous examples that despite the well-known mathematical issues of the OEP with finite basis 
sets, our OEP equations are numerically robust. 
\color{black} 
We demonstrate that appropriately constraining the `screening charge' which corresponds to the Hartree, exchange and correlation potential not only corrects its 
asymptotic behaviour but also allows the exchange and correlation potential to exhibit a nonzero derivative discontinuity, a feature of the exact KS potential that is 
necessary for the accurate prediction of band-gaps in solids but very hard to capture with semi-local approximations.
\end{abstract}

\maketitle

\section{Introduction}

A challenge with common density functional approximations is the imbalance of accuracy between the energy functionals and the corresponding Kohn-Sham (KS)
potentials, i.e. the functional derivatives of the energy density-functionals. Although the accuracy and quality of an energy density-functional is often quite good, 
the resulting KS potential is inferior \cite{adventures_bartlett,inconsistent,kieron_density_error}. 
The quest to derive ever more accurate energy density-functionals to obtain moderate improvements on the KS potential may not be the best strategy (it is vulnerable to diminishing returns in the accuracy of the KS potential). We explore different routes to improved accuracy for these calculations. 
 
Previously, we explored the minimisation of potential functionals defined by an energy difference, instead of density fucntionals of the total energy, 
as a means of improving the quality of the KS potential
\cite{gidopoulos_ks,lfx,callow_gidopoulos,helgaker_teale}. 
The advantage of this approach is that the energy difference is bound from below, even in approximations from finite-order (second) perturbation theory; the latter can then be employed directly to derive accurate exchange-correlation (xc) potentials without the risk of variational collapse \cite{gidopoulos_ks,callow_gidopoulos}.

In this paper, we review briefly and expand on our work \cite{clda,clda_review,cdfa} to improve the performance of local and semi-local density-functional 
approximations (DFAs), by imposing physical constraints on the single-particle, local, effective (KS) potential, whose orbitals minimise the total energy functional. 
In Refs. \onlinecite{clda,clda_review,cdfa} we argued that these constraints improve the asymptotic behaviour and overall quality of the KS 
potential by removing the erroneous effects of self-interactions (SIs). As evidence, we demonstrated that, 
compared with the results from the unconstrained minimisation, the ionisation potentials (IPs) 
\footnote{Calculated as the negative of the HOMO eigenvalue.} of a large number of atoms, molecules, even anions, obtained from our constrained
 minimisation improved significantly, while the calculated total energies increased only minimally. 

In this work, we further show that with a judicious choice, the constraints imposed on the KS potential of local and semi-local DFAs enable their 
(constrained) exchange and correlation (xc) potential to exhibit exotic, non-analytic behaviour, 
expected only in more elaborate and computationally costly levels of theory, or from higher, heavenly rungs on Jacob's ladder of DFAs, 
as envisaged by John Perdew and co-workers \cite{jacob_ladder}. 
   
\section{Constrained minimisation of density functional approximations} \label{sec:ii}

\color{black}
In the constrained minimisation method \cite{clda,clda_review,cdfa} we employ the standard total energy expression in DFT, 
using a density functional approximation (DFA) for the xc energy density-functional, $E_\textrm{xc}^\textrm{DFA}[\rho]$,
\begin{equation}\label{eq:E_tot_KS}
    E_{v_{\rm en}}^{\rm DFA} [\rho] = T_\textrm{s}[\rho] + \int \dd{\rr} v_\textrm{en}(\rr) \rho(\rr) + U [\rho] + E_\textrm{xc}^\textrm{DFA}[\rho] .
\end{equation}
The various quantities have their usual definitions, $v_{\rm en}$ is the external potential, $T_s [ \rho ]$, $U [ \rho ] $ are the noninteracting kinetic energy and Hartree energy density functionals.
Following the optimised effective potential method (OEP) \cite{sharp,talman}, we set that the KS orbitals satisfy single-particle KS equations employing an effective potential $v ( \rr )$,
\begin{equation}
 \label{ks_eq}
\bigg[ - { \nabla^2 \over 2} + v_\textrm{en} ( \rr ) + v ( \rr )  \bigg] \phi_i ( \rr ) = \epsilon_i \, \phi_i ( \rr ) .
\end{equation}
The total energy is then minimised by imposing constraints, detailed below, on the effective potential $v(\rr)$.
The effective potential $v ( \rr )$ is akin to the Hartree-exchange and correlation (Hxc) potential of KS theory $v_{\rm Hxc}^{\rm DFA} (\rr) $. 
However, the constraints we impose correct the asymptotic form of $v(\rr)$ and alleviate other effects of SIs from it, so in general, $v (\rr) \ne v_{\rm Hxc}^{\rm DFA} (\rr)$.

In the constrained method, we treat the Hxc screening density, or electron repulsion density, $\rho_{\rm scr} (\rr)$,
\footnote{For brevity we often just use the term `screening density'. It should not be confused with the similar term by Baerends et al.},
as the fundamental quantity. It is defined via Poisson's equation from the Laplacian of the difference of the (exact or approximate) 
KS potential minus the external potential, $\nabla^2 \big[ v_s ( \rr ) - v_{\rm en} ( \rr ) \big]$; 
for example, the Hxc screening density of the exact KS potential is given by,
\begin{equation} \label{eq3}
    \rho_\textrm{scr}(\rr) = -\frac{1}{4\pi} \nabla^2 v_\textrm{Hxc}(\rr) .
\end{equation}
Together with the integrated Hxc screening charge $Q_\textrm{scr}$,
\begin{equation} 
    Q_\textrm{scr}  = \int \dd{\vec{r}} \rho_\textrm{scr} (\rr),
\end{equation}
the Hxc screening density plays a central role in our constrained method to mitigate against the effects of self-interactions. 
The concept of an effective screening density was first explored 
in Refs.~\onlinecite{gorling_poisson,liu_ayers_parr,ayers_levy} in terms of a screening density for the xc (or exchange only) potential; for the exact xc-potential, 
the screening charge is $Q_\textrm{scr}^\textrm{xc}=-1$ \cite{gorling_poisson,liu_ayers_parr,ayers_levy}. 
It has been used in various applications of the OEP method to fix the freedom of a constant in the 
OEP solution \cite{Hirata_oep,gorling_poisson,Gorling_numerically_stable_oep,Gorling_balanced_2,ROHR2006193}.

The Hxc-screening charge of the exact KS potential satisfies the intuitive sum rule \cite{gorling_poisson,liu_ayers_parr,ayers_levy},
\begin{equation} \label{eq:sum_rule}
    Q_\textrm{scr} =\int\dd{\vec{r}} \rho_\textrm{scr} (\vec{r}) = N-1 .
\end{equation}
However, in common DFAs (such as the local density approximation, L(S)DA, and most generalized-gradient approximations (GGAs)) this sum rule is violated and the screening charge 
is in fact given by $Q_\textrm{scr}=N$. We argue \cite{clda} that this violation of the sum rule can be attributed to the presence of SIs, since it implies that 
any of the electrons of an $N$-electron system are effectively repelled, via the Hxc potential, by a net charge of $N$ electrons. We note that the sum 
rule \eqref{eq:sum_rule}, which depends on the screening density and is violated for LDA and common GGAs, is different from the well-known sum rule  \cite{gross_book} for the xc hole, 
$\int d \rr ' \, \rho_{\rm xc} ( \rr , \rr ' ) = -1$, which is satisfied by LDA and common GGAs. 
The quantities $\rho_\textrm{scr}^\textrm{xc}(\vec{r})$ and $\rho_\textrm{xc}(\vec{r},\vec{r'})$ are not directly related.

Accordingly, in the constrained minimisation of DFAs \cite{clda,clda_review,cdfa} (which we henceforth refer to as the CDFA method), our strategy to mitigate 
the effects of SIs from the effective potential is to enforce that the KS orbitals satisfy Eq.~\eqref{ks_eq} with the effective potential 
$v(\rr)$ represented by the effective screening density
\begin{equation} \label{v_rho}
v ( \rr ) = \int \dd{\rr'} \ { \rho_\textrm{scr} (\rr ') \over | \rr - \rr ' | } ,
\end{equation}
where $\rho_\textrm{scr} ( \rr )$ satisfies two constraints:
\begin{equation} \label{norm}
Q_\textrm{scr} = N-1 ,
\end{equation}
and
\begin{equation} \label{pos}
\rho_\textrm{scr} ( \rr ) \ge 0 .
\end{equation}
The second constraint \eqref{pos} is physically intuitive, hinting at interpreting  $\rho_\textrm{scr} ( \rr ) $ as the charge density of $N-1$ electrons. 
However, this condition is too restrictive and not satisfied by the exact KS potential.

Nonetheless, the positivity constraint \eqref{pos} has a double role in the constrained minimisation method. As explained in Refs.~\onlinecite{clda,clda_review,cdfa}, 
the CDFA minimization procedure must be solved within the optimized effective potential (OEP) framework \cite{sharp,talman}. 
Primarily, the positivity constraint allows the mathematical problem of constrained minimisation to 
remain well posed in the limit of complete orbital and auxiliary basis sets \cite{clda,clda_review}; without the positivity constraint, 
there is nothing to prevent the screening density from separating into a component in the energetically important spatial region near the molecule, 
with charge $Q^a_\textrm{scr}=N$, and a separate component with charge $Q^b_\textrm{scr}=-1$ pushed out to infinity (within the basis set limits). 
Secondly, the solution of the OEP equation in Gaussian basis set codes is a longstanding problem in DFT, which has hindered the widespread 
adoption of OEP-based methods in practical calculations. Various methods have been developed to overcome these numerical difficulties 
which typically manifest themselves as spurious oscillations in the effective potential \cite{Hirata_oep,Stavoverov_OEP,Gorling_numerically_stable_oep}.
With finite orbital and auxiliary basis sets, the positivity constraint \eqref{pos} offers a simple way to reduce drastically the variational 
flexibility of $\rho_\textrm{scr} ( \rr ) $ and of $v ( \rr )$ and thereby helps to overcome mathematical pathologies
in the solution of the OEP equation.

In the previous implementation of the CDFA method, the positivity constraint was used in combination with a singular value decomposition 
(SVD) of the density-density response matrix to ensure the solution of the OEP equation is well-behaved. Instead, here we apply the method of 
Ref.~\onlinecite{nonanalycity_oep} to solve the OEP equation in the CDFA method. 
We review the main ideas below; see Ref.~\onlinecite{nonanalycity_oep} and the subsequent discussion 
in Refs.~\onlinecite{comment} and \onlinecite{reply} for details.

The OEP equation (Fredholm integral equation of the first kind) is obtained by taking the functional derivative of an energy term with respect to the density 
(e.g. $T_s [ \rho]$, $E_{\rm x} [\rho]$, $E_{\rm xc} [ \rho]$) 
when this energy term is written as an implicit functional of the density. Alternatively, it can be obtained by minimising the DFT total energy expression,
\eqref{eq:E_tot_KS},  indirectly by searching for the effective potential $v(\rr)$ in \eqref{ks_eq} 
whose KS orbitals minimise the total energy \cite{grabo1997,engel2003orbital}. Either way, we obtain the integral OEP equation,      
\begin{equation} \label{fredholm}
    \int \dd{\rr'} \chi_v(\rr,\rr') \, v (\rr) = b_v(\rr),
\end{equation}
where $ \chi_v (\rr,\rr')$ is the density-density response function given by (in a complete orbital basis set)
\begin{equation}\label{eq:oep_general}
    \chi_v ( \rr , \rr ' ) = 2\sum_i^{\rm occ} \sum_{a}^{\rm unocc} { \phi_i ( \rr ) \, \phi_a ( \rr ) \, \phi_i ( \rr ' ) \, \phi_a ( \rr ' )  
\over \epsilon_i - \epsilon_a} .
\end{equation}
The KS orbitals from \eqref{ks_eq} are assumed to be real-valued. 
The right-hand side (RHS) $b_v(\rr)$ depends on the energy term whose functional derivative we take, in our case the Hxc energy $U[\rho] + E_{\rm xc}^{\rm DFA} [ \rho]$. It is given by
\begin{equation}
    b_v(\rr) = 2\sum_i^{\rm occ} \sum_{a}^{\rm unocc} \frac{\mel{\phi_i}{v_\textrm{H}+ \fdv{E_\textrm{xc}^\textrm{DFA}}{\rho}}{\phi_a}}{\epsilon_i - \epsilon_a} \phi_i(\rr) \phi_a(\rr).
\end{equation}
If no constraints are imposed, the solution of \eqref{fredholm} is trivially $v ( \rr ) = v_{\rm Hxc}^{DFA} ( \rr )$ within a constant, since $ \chi_v ( \rr , \rr ' )$ has no null eigenfunctions except the constant function.   
In Ref.~\onlinecite{cdfa} we explain how we impose the normalisation constraint (\ref{norm}) on the effective screening density and demonstrate that the 
the scheme can be applied for any given DFA, including LDA, GGAs and hybrid functionals.

To understand the effect of finite orbital basis sets on the solution of the OEP equation, we focus on the density-density response function; 
the analysis below also applies to the RHS $b_v(\rr)$. We split $\chi_v$ into two terms, the first of which can be represented exactly in the orbital basis, and the 
second which must be approximated. $\chi_v$ is given, for $\lambda=1$, by
\begin{equation} \label{chi_lambda}
\chi_v^\lambda ( \rr , \rr ' ) = \chi_v^0 ( \rr , \rr ' ) + \lambda \, \bar \chi_v ( \rr , \rr ' ) ,
\end{equation}
with
\begin{equation}
\chi_v^0 ( \rr , \rr ' ) = 2\sum_i^{\rm occ} \sum_{a \in {\rm OB}}^{\rm unocc} { \phi_i ( \rr ) \, \phi_a ( \rr ) \, \phi_i ( \rr ' ) \, \phi_a ( \rr ' )  
\over \epsilon_i - \epsilon_a},
\end{equation}
\begin{equation} \label{bar_chi}
\bar \chi_v ( \rr , \rr ' ) = 2\sum_i^{\rm occ} \sum_{b \notin {\rm OB}}^{\rm unocc} { \phi_i ( \rr ) \, \phi_b ( \rr ) \, \phi_i ( \rr ' ) \, \phi_b ( \rr ' )  
\over \epsilon_i - \epsilon_b}.
\end{equation}
The sum is over occupied $\{ \phi_i \}$ and unoccupied $\{ \phi_a , \phi_b\}$ KS orbitals \eqref{ks_eq} in the KS Slater determinant. 
We assume for simplicity that the orbital basis set (OB) is composed exactly of a set of low lying KS orbitals, ${\rm OB} = \{ \phi_i \} \cup \{ \phi_a \}$, 
i.e., the set of orbitals which are occupied in the KS state and the lowest unoccupied ones.
Until Ref.~\onlinecite{nonanalycity_oep}, when working with finite orbital basis sets, the second part $\bar \chi_v$ of the response function, which we denote the `complement' of the response function, was typically omitted. 

By definition, the complement $\bar{\chi}_v$ cannot be represented exactly so we  must approximate it. We use the \"{U}nsold approximation \cite{unsold} together with the completeness relation for the KS orbitals (in much the same manner as the well-known Krieger-Li-Iafrate (KLI) approximation 
\cite{kli1,kli2} and common energy demoninator approximation (CEDA) \cite{CEDA,localizedHF} methods), in which case  $\bar{\chi}_v$ reduces to
\begin{equation}\label{bar_chi_approx}
    \bar{\chi}_v(\rr,\rr')=-\frac{2}{\Delta} \sum_i^\textrm{occ} \orb{i}{r} \orb{i}{r'} \left\{ \delta(\rr-\rr') -\sum_j^{\textrm{occ}} \orb{j}{r}\orb{j}{r'} - \cancel{\sum_{a \in {\rm OB}}^{\rm unocc} \orb{a}{r}\orb{a}{r'}} \right\},
\end{equation}
where $- \Delta$ is the common energy denominator that replaces $\epsilon_i - \epsilon_b$ in \eqref{bar_chi}, $\Delta > 0$.
In Eq.~\eqref{bar_chi_approx}, we omit the final term with the same domain as $\chi_v^0$, because its contribution to $\chi_v^\lambda$ vanishes for small $\lambda$, 
which is ultimately the limit we seek.

We observe that, as long as $\Delta > 0$, the value of $\Delta$ does not play a role in the results, 
since $\Delta$ always appears together with $\lambda$, in the ratio $\lambda / \Delta$, 
and we investigate the limit $\lambda \rightarrow 0$.
We shall also consider the limit $\lambda \rightarrow \infty$, for which the value of positive $\Delta$ does not matter either. It is straightforward to confirm that $\bar \chi_v$ is negative semi-definite, like $\chi_v^0 $, and that the only 
null eigenfunction of $\bar \chi_v$ is the constant function. 

The same procedure is applied for the RHS $b_v(\rr)$ of the OEP equation \eqref{eq:oep_general}, which yields the following expressions for the terms $b_v^0(\rr)$ and its complement $\bar{b}_v(\rr)$,
\begin{align} \label{eq:b0}
     b_v^0(\rr) &= 2\sum_i^{\rm occ} \sum_{a}^{\rm unocc} \frac{\mel{\phi_i}{v_\textrm{H}+ \fdv{E_\textrm{xc}^\textrm{DFA}}{\rho}}{\phi_a}}{\epsilon_i - \epsilon_a} \phi_i(\rr) \phi_a(\rr) \\
     \nonumber
     \bar{b}_v(\rr) &= -\frac{2}{\Delta} \sum_{i=1}^\textrm{occ} \Bigg\{\phi_i(\rr) \int \dd{\rr'} \delta(\rr-\rr') \left( v_\textrm{H}(\rr')+ \fdv{E_\textrm{xc}^\textrm{DFA}}{\rho(\rr')}  \right) \phi_i(\rr') \\
     & \hspace{5em} - \sum_{j}^{\rm occ} \mel**{\phi_i}{v_\textrm{H}+ \fdv{E_\textrm{xc}^\textrm{DFA}}{\rho}}{\phi_j} \phi_i(\rr) \phi_j(\rr) \Bigg\}.
\end{align}
The OEP equation thus takes the following form,
\begin{equation} \label{oep_l}
\int\dd{\rr'} \big[ \chi_v^0 ( \rr , \rr ' ) +  \lambda  \, \bar \chi_v ( \rr , \rr ' ) \big] v^ \lambda  ( \rr' ) = b_v^0 ( \rr ) +  \lambda  \, \bar b_v ( \rr ).
\end{equation}
To solve this equation in a Gaussian basis set code, the screening density is expanded in an auxiliary basis set and its coefficients can be found by a straightforward matrix inversion. 
The screening charge constraint \eqref{norm} besides mitigating against SI errors is also necessary to fix the freedom of a constant in the effective potential \cite{Hirata_oep} and 
is enforced using a Lagrange multiplier. The optimization procedure is explained in detail in Ref.~\onlinecite{cdfa}; the only difference here is that the matrices for the LHS and RHS of 
the OEP equation now contain the additional complement terms.

\color{black}
Prior to Ref.~\onlinecite{nonanalycity_oep}, the finite orbital basis OEP was given by the solution of \eqref{oep_l} at $\lambda  = 0$.
However, this solution 
leaves the effective potential $v^0 ( \rr ) $ indeterminate in the null space of $\chi_v^0$, which is infinite-dimensional. 
In order to obtain a smooth potential, $v^0 ( \rr )$, one must restrict the freedom of $v^0 ( \rr )$, which has spawned a variety of approaches in the literature. These include, for example, schemes to balance the relative sizes of the orbital and auxiliary basis set \cite{Gorling_numerically_stable_oep,Gorling_balanced_2}; regularization techniques to smooth the effective potential \cite{Yang_regularized_1,Yang_regularized_2}; and removing the additional freedom of an auxiliary basis set \cite{Kollmar_Filatov_OEP}. In our method, rather than restricting the freedom of $v^0 ( \rr )$, we instead solve the OEP equation to find the potential $v^\lambda(\rr)$ which is defined mathematically to be unique for finite $\lambda$.
 
The main point of Ref.~\onlinecite{nonanalycity_oep} is the observation that 
the solution of the same equation \eqref{oep_l} for any finite $ \lambda  > 0$, even $ \lambda  $ tending to zero, 
determines the effective potential fully, up to a constant. 
The extension of the response function with $\bar \chi_v$ amounts to using an effectively complete orbital basis.
Numerically, we find that the solution of \eqref{oep_l} is smooth for almost any $ \lambda  > 0$ 
\footnote{There are some restrictions for this to always hold: (i) the auxiliary basis cannot be significantly larger than the orbital basis and (ii) the value of $\lambda$ 
cannot be arbitrarily small in numerical applications}, 
including the limits for small and for large $ \lambda $, which correspond respectively to the OEP potential in a finite orbital basis, $v^{ \lambda  \rightarrow 0} ( \rr ) $, 
and to its (Uns\"old) approximation with a common energy denominator, $v^{\infty} ( \rr ) $. It turns out that for the effective xc potentials in the constrained minimisation method, 
the two solutions are close to each other.

\subsection{Relaxing the positivity constraint}\label{sec:2a}

In Refs.~\onlinecite{clda,clda_review,cdfa} we solved the OEP equation for CDFA method, 
using finite orbital and auxiliary basis sets, with $ \lambda =0$.
The indeterminacy of the effective potential was restricted by expressing $v(\rr)$ in terms of the screening density $\rho_\textrm{scr} ( \rr ) $ 
in \eqref{v_rho} and then constraining the screening charge $Q_\textrm{scr}$ \eqref{norm} as well as the sign of $\rho_\textrm{scr} ( \rr )$ \eqref{pos}. 

However, the positivity constraint, implemented with a penalty function \cite{cdfa} is a computational bottleneck for the calculation.
In a forthcoming paper, we implement the positivity constraint more efficiently, by writing $\rho_\textrm{scr} (\rr) = | f_\textrm{scr} ( \rr ) |^2 $, and
solving for the screening amplitude $f_\textrm{scr} ( \rr ) $ \cite{fscr}, which ensures the constrained minimization is mathematically well-posed regardless of basis set size.

In the next part, we investigate the effects of relaxing the positivity constraint on the convergence of the screening potential
and screening density.
A weak effect, for sufficiently flexible auxiliary basis sets, will justify the relaxation of the 
positivity constraint and reduce the computational effort.
The auxiliary basis sets we use are un-contracted cc-pV$X$Z \cite{Dunning_1,Dunning_2}, with $X$=D,T,Q.


In the rest of the section, we show indicative results for the CDFA method applied to the LDA functional, henceforth denoted by CLDA, 
where the minimisation was performed under just 
the constraint for the screening charge, $Q_\textrm{scr} = N-1$ \eqref{norm}.  
In order to determine $v(\rr )$ and $\rho_\textrm{scr} ( \rr )$, we employ the extended response function $\chi_v^ \lambda  ( \rr , \rr ' )$, in the limit of small
$ \lambda $. We use $\lambda/\Delta = 0.01$, but the results seem converged and do not change if we reduce $\lambda / \Delta$ by an order of magnitude.
The positivity constraint enabled the constrained minimisation problem to remain well posed in the limit of large
(complete) orbital and auxiliary basis sets. Consequently, we expect the screening charge to change gradually, as we increase the size 
of the auxiliary basis. This effect will be stronger for systems with few electrons, since then, the difference between $N-1$ and $N$ is largest.

Calculations were performed in the Gaussian basis set code HIPPO \footnote{Contact NNL at lathiot@eie.gr for information}, with one- and two-electron integrals for the Cartesian Gaussian basis elements 
calculated 
using the GAMESS code \cite{gamess1,gamess2}. Basis set data was obtained from the Basis Set Exchange database \cite{bse}.

\begin{figure}[h!]
  \centering
  \subfloat[Aux basis unc.~cc-pVDZ]
  {\includegraphics[trim=0 0 0 0 , clip, width=0.33\textwidth]{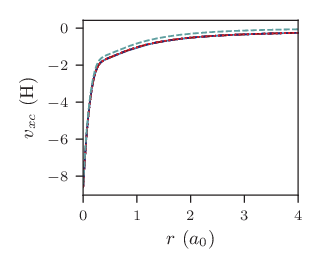}\label{fig:1a}}
  \subfloat[Aux basis unc.~cc-pVTZ]
  {\includegraphics[trim=0 0 0 0 , clip, width=0.33\textwidth]{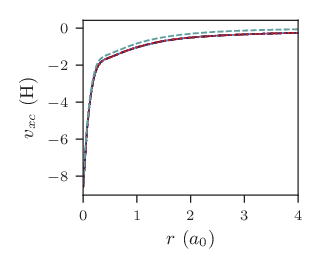}\label{fig:1b}}
  \subfloat[Aux basis unc.~cc-pVQZ]
  {\includegraphics[trim=0 0 0 0 , clip, width=0.33\textwidth]{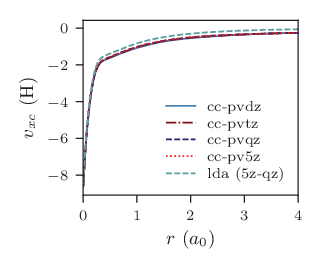}\label{fig:1c}}
   \caption{Ne atom, CLDA xc potentials $v_\textrm{xc} ( r ) $ using fixed auxiliary basis sets with various orbital basis sets. Blue dashed line is LDA. 
   }
\end{figure}
\begin{figure}[h!]
  \centering
  \subfloat[Aux basis unc.~cc-pVDZ]
  {\includegraphics[trim=0 0 0 0 , clip, width=0.33\textwidth]{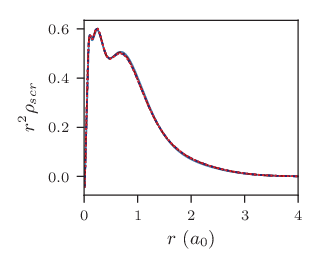}\label{fig:2a}}
  \subfloat[Aux basis unc.~cc-pVTZ]
  {\includegraphics[trim=0 0 0 0 , clip, width=0.33\textwidth]{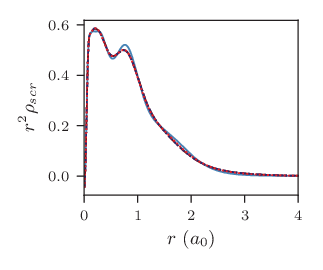}\label{fig:2b}}
  \subfloat[Aux basis unc.~cc-pVQZ]
  {\includegraphics[trim=0 0 0 0 , clip, width=0.33\textwidth]{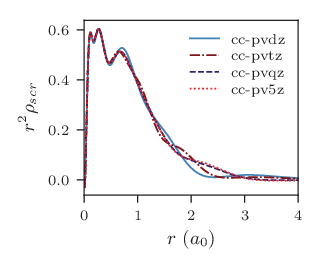}\label{fig:2c}}
   \caption{Ne atom, CLDA results for $r^2 \rho_\textrm{scr} ( \rr ) $. Fixed auxiliary basis set for the expansion of $\rho_\textrm{scr} ( r ) $ 
   in each subfigure, various orbital basis sets. 
   }
\end{figure}
In Figs.~\ref{fig:1a}-\ref{fig:1c}, the CLDA xc potential is shown for the Ne atom and three auxiliary basis sets, un-contracted cc-pV$X$Z, with $X$=D,T,Q.  
In each sub-figure $v_\textrm{xc}^\textrm{CLDA} ( r) $ is shown for fixed auxiliary basis and 
various orbital basis sets: cc-pV$X$Z, with $X$=D,T,Q,5. For comparison, the LDA potential $v_\textrm{xc}^\textrm{LDA} ( r )$
is also shown with a blue dashed line. 

In Figs.~\ref{fig:2a}-\ref{fig:2c}, $r^2 \rho_\textrm{scr} ( r ) $ (CLDA screening density multiplied by $r^2$), 
is shown for the Ne atom and three auxiliary basis sets, un-contracted cc-pV$X$Z, with $X$=D,T,Q.  
In each sub-figure $r^2 \rho_\textrm{scr} ( r ) $ is shown for fixed auxiliary basis and various orbital basis sets: cc-pV$X$Z, with $X$=D,T,Q,5. 
The overall convergence of the xc potential is excellent. The convergence of $\rho_{scr} ( r ) $ for fixed aux basis and increasing size of orbital basis is also 
very good. 
Before proceeding, it is worth noting that despite not deploying the positivity constraint \eqref{pos} that would restrict 
the flexibility of the screening density and the xc potential, the latter (solutions of CLDA-OEP equations (12) and (15) in Ref.~\onlinecite{cdfa}) 
turn out to be smooth functions, not showing any wild oscillations characteristic of OEP-finite-basis pathologies, 
for any combination of orbital and auxiliary basis sets. This confirms our claim that by extending the domain of the 
density-density response function (\ref{chi_lambda}, \ref{oep_l}), the solution of finite-basis-OEP equations is well behaved.

\begin{figure}[h!]
  \centering
  \subfloat[Aux basis unc.~cc-pVDZ]
  {\includegraphics[trim=0 0 0 0 , clip, width=0.33\textwidth]{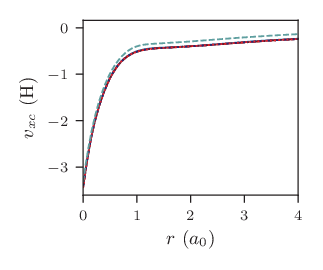}\label{fig:3a}}
  \subfloat[Aux basis unc.~cc-pVTZ]
  {\includegraphics[trim=0 0 0 0 , clip, width=0.33\textwidth]{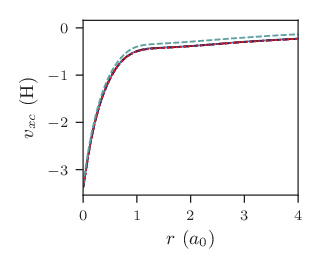}\label{fig:3b}}
  \subfloat[Aux basis unc.~cc-pVQZ]
  {\includegraphics[trim=0 0 0 0 , clip, width=0.33\textwidth]{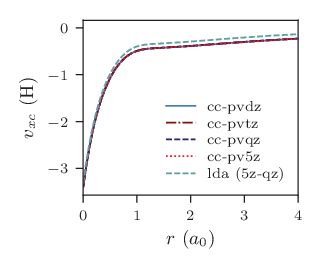}\label{fig:3c}}
   \caption{Be atom, CLDA xc potentials $v_\textrm{xc} ( r ) $ using fixed auxiliary basis sets with various orbital basis sets. Blue dashed line is LDA. }
\end{figure}
\begin{figure}[h!]
  \centering
  \subfloat[Aux basis unc.~cc-pVDZ]
  {\includegraphics[trim=0 0 0 0 , clip, width=0.33\textwidth]{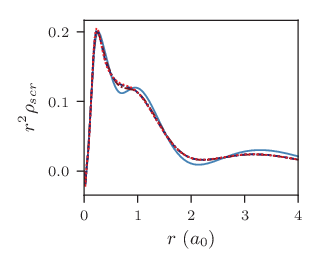}\label{fig:4a}}
  \subfloat[Aux basis unc.~cc-pVTZ]
  {\includegraphics[trim=0 0 0 0 , clip, width=0.33\textwidth]{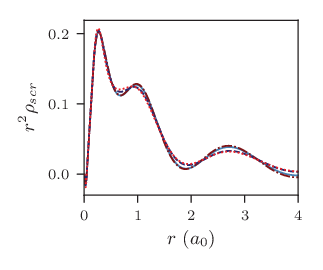}\label{fig:4b}}
  \subfloat[Aux basis unc.~cc-pVQZ]
  {\includegraphics[trim=0 0 0 0 , clip, width=0.33\textwidth]{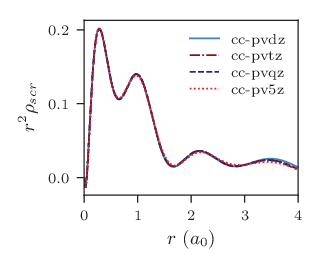}\label{fig:4c}}
   \caption{Be atom, CLDA results for $r^2 \rho_\textrm{scr} ( \rr ) $. Fixed auxiliary basis set for the expansion of $\rho_\textrm{scr} ( r ) $ 
   in each subfigure, various orbital basis sets. }
\end{figure}

\noindent 
Figs.~\ref{fig:3a}-\ref{fig:3c}, \ref{fig:4a}-\ref{fig:4c} show similar results to previous Figs.~\ref{fig:1a}-\ref{fig:1c}, \ref{fig:2a}-\ref{fig:2c}, 
but for the Be atom.

\begin{figure}[h!]
  \centering
  \subfloat[Aux basis unc.~cc-pVDZ]
  {\includegraphics[trim=0 0 0 0 , clip, width=0.33\textwidth]{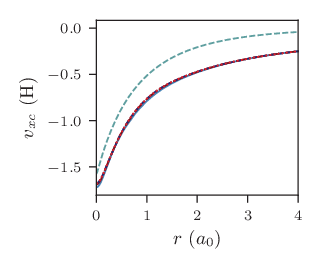}\label{fig:5a}}
  \subfloat[Aux basis unc.~cc-pVTZ]
  {\includegraphics[trim=0 0 0 0 , clip, width=0.33\textwidth]{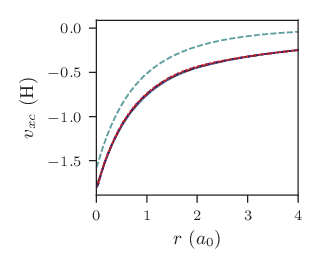}\label{fig:5b}}
  \subfloat[Aux basis unc.~cc-pVQZ]
  {\includegraphics[trim=0 0 0 0 , clip, width=0.33\textwidth]{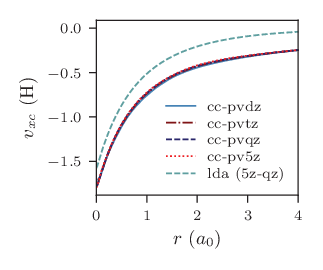}\label{fig:5c}}
   \caption{He atom, CLDA xc potentials $v_\textrm{xc} ( r ) $ using fixed auxiliary basis sets with various orbital basis sets. Blue dashed line is LDA. 
}
\end{figure}
\begin{figure}[h!]
  \centering
  \subfloat[Aux basis unc.~cc-pVDZ]
  {\includegraphics[trim=0 0 0 0 , clip, width=0.33\textwidth]{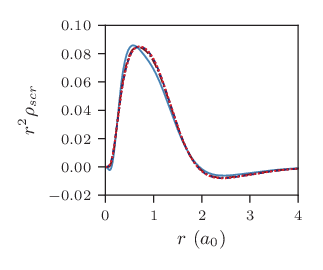}\label{fig:6a}}
  \subfloat[Aux basis unc.~cc-pVTZ]
  {\includegraphics[trim=0 0 0 0 , clip, width=0.33\textwidth]{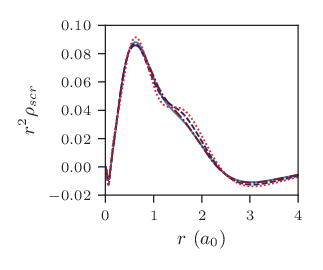}\label{fig:6b}}
  \subfloat[Aux basis unc.~cc-pVQZ]
  {\includegraphics[trim=0 0 0 0 , clip, width=0.33\textwidth]{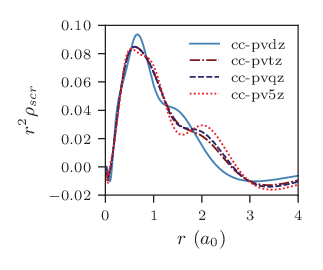}\label{fig:6c}}
   \caption{He atom, CLDA screening densities $\rho_\textrm{scr} ( \rr ) $ expanded in fixed auxiliary basis sets with various orbital basis sets. 
}
   \label{fig:Her}
\end{figure}
\noindent 
We proceed to discuss Figs.~\ref{fig:5a}-\ref{fig:5c}, \ref{fig:6a}-\ref{fig:6c}, which show similar results as Figs.~\ref{fig:1a}-\ref{fig:1c}, \ref{fig:2a}-\ref{fig:2c}, 
and \ref{fig:3a}-\ref{fig:3c}, \ref{fig:4a}-\ref{fig:4c}, for the He atom. The convergence of the xc potential 
{\em for fixed auxiliary basis} and increasing orbital basis size is good. Note that for any combination of 
orbital and auxiliary basis, the xc potential is smooth.
The convergence of the screening density {\em for fixed auxiliary basis} and increasing size of orbital basis is slower than the other systems. 
In addition, as the size of the auxiliary basis increases, from \ref{fig:6a} to \ref{fig:6b} to \ref{fig:6c}, 
the screening density keeps changing considerably.
Note specifically the negative part of the screening density in Figs.~\ref{fig:6a}-\ref{fig:6c}.  In Fig.~\ref{fig:6a} the negative lump is centred around 
2.5 $a_0$ away from the origin, in Fig.~\ref{fig:6b} it is centred around 3.0 $a_0$ away from the origin and in Fig.~\ref{fig:6c} it has moved to 3.5 $a_0$.
This is the effect we discussed in section \ref{sec:ii}. The positivity constraint enables the constrained minimisation problem to remain well posed 
for large basis sets (here large auxiliary bases). 
With only the constraint on $Q_\textrm{scr}$ enabled and without positivity, it becomes energetically preferable, during the total energy minimisation, 
to converge to a screening density with the screening charge locally equal to $N$ (=$Q_\textrm{scr}^\textrm{LDA}$), 
and to shift negative charge density away from the system. 
The effect is already evident for the moderately large auxiliary bases used in our study, because the difference between $N-1$ and $N$ is relatively 
large for $N=2$.

\begin{figure}[h!]
  \centering
  \subfloat[Aux basis unc.~cc-pVDZ]
  {\includegraphics[trim=0 0 0 0 , clip, width=0.33\textwidth]{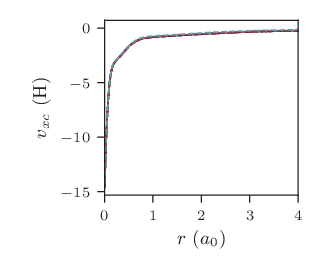}\label{fig:7a}}
  \subfloat[Aux basis unc.~cc-pVTZ]
  {\includegraphics[trim=0 0 0 0 , clip, width=0.33\textwidth]{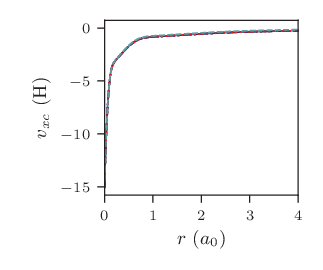}\label{fig:7b}}
  \subfloat[Aux basis unc.~cc-pVQZ]
  {\includegraphics[trim=0 0 0 0 , clip, width=0.33\textwidth]{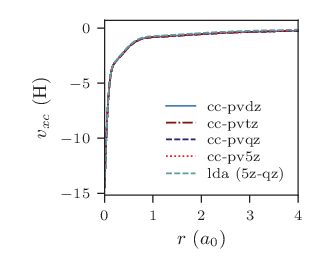}\label{fig:7c}}
   \caption{Cl$^-$ anion, CLDA xc potentials $v_{xc} ( r ) $ using fixed auxiliary basis sets with various orbital basis sets (augmented). 
   Blue dashed line is LDA. Convergence with increasing size of orbital basis is evident.}
\end{figure}
\begin{figure}[h!]
  \centering
  \subfloat[Aux basis unc.~cc-pVDZ]
  {\includegraphics[trim=0 0 0 0 , clip, width=0.33\textwidth]{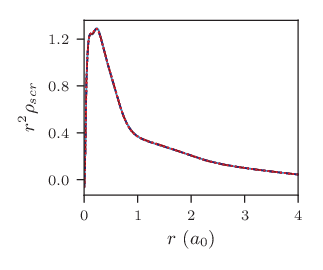}\label{fig:8a}}
  \subfloat[Aux basis unc.~cc-pVTZ]
  {\includegraphics[trim=0 0 0 0 , clip, width=0.33\textwidth]{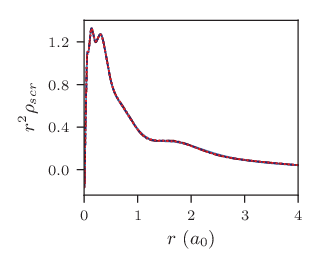}\label{fig:8b}}
  \subfloat[Aux basis unc.~cc-pVQZ]
  {\includegraphics[trim=0 0 0 0 , clip, width=0.33\textwidth]{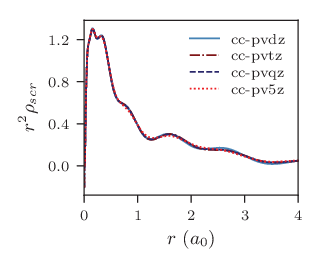}\label{fig:8c}}
   \caption{Cl$^-$ anion, CLDA screening densities $\rho_\textrm{scr} ( \rr ) $ expanded in fixed auxiliary basis sets with various orbital basis sets.} 
   \label{fig:Clr1}
\end{figure}

The negatively charged ions is another class of difficult systems where LDA fails qualitatively.
In Figs.~\ref{fig:7a}-\ref{fig:7c}, \ref{fig:8a}-\ref{fig:8c} we plot the CLDA xc potential and screening density of the chlorine anion Cl$^-$. 
The orbital basis sets are augmented cc-pV$X$Z, with $X$=D,T,Q,5. It is evident that both the CLDA xc potential and the CLDA 
screening density are well converged and these systems do not present a challenge to the constrained minimisation, at least regarding convergence.

In Table~\ref{table:t} we show the IPs of several systems, including anions, obtained as the negative of the HOMO eigenvalue.
For comparison with our previous CLDA method, in which we had imposed the positivity constraint, we show the CLDA IPs with (fourth column) and without
positivity (fifth column). The results with positivity are from Ref.~\onlinecite{clda}. The resulting IPs do not depend strongly on the 
positivity constraint, except in helium, where we see a larger difference. We still see the familiar improvement of CLDA over the LDA results.
\begin{table}[h!]
\begin{tabular}{ | c  | c |  c | c | c || c |}
 \hline
 & Basis &  LDA & CLDA pos & CLDA no pos & Exp \\
\hline
\hline
He        &  T-Q  &  15.46 & 23.14& 21.57 &  24.6   \\
Be        &  T-T  &   5.59 & 8.62 &  8.11 & 9.32   \\
Ne        &  T-T  &   13.16&  18.94 & 18.94 &  21.6   \\
H$_2$O    &  T-T     & 6.96&   11.24 & 11.34 & 12.8   \\
NH$_3$    &  T-T     & 6.00&   9.81 & 9.77  & 10.8  \\
CH$_4$    &  D-D    & 9.28&  12.52  & 10.51  & 14.4  \\
C$_2$H$_2$&  D-D     & 7.02&   10.63 & 10.31 &  11.5  \\
C$_2$H$_4$&  D-D    & 6.67&   9.57  & 9.35 & 10.7  \\
CO        &  D-D  &    8.75&   12.73 & 12.11 & 14.1 \\
NaCl      &  D-D  &  5.13&   7.87  & 7.82 & 8.93 \\
\hline\hline
F$^-$&T\footnotemark[1]-T   & $\epsilon_{\rm H}>0$  & 2.23 & 2.16 & 3.34 \\
Cl$^-$&T\footnotemark[1]-T &  $\epsilon_{\rm H}>0$  & 2.61 & 2.59 & 3.61\\
OH$^-$&T\footnotemark[1]-T &  $\epsilon_{\rm H}>0$  & 0.99 & 0.93 & 1.83\\
CN$^-$&T\footnotemark[1]-T       &   0.13 & 2.87 & 2.86 & 3.77 \\
\hline
\hline
\end{tabular}
\caption{ 
The IPs of selected atoms, molecules (top) and negative ions (bottom) are shown in columns 3-5 . 
The IPs are obtained as the negative of the HOMO eigenvalue $\epsilon_{\rm H}$ of the neutral system or the anion. 
The positivity constraint is employed for the results in column 4 (from Ref.~\onlinecite{clda}) and relaxed for the results in column 5.  
The experimental IPs and electron affinities are shown in the sixth column.
In the second column, $X$-$Y$ stands for basis sets cc-pV$X$Z and un-contracted cc-pV$Y$Z for the expansion of orbitals and screening charge densities.
All energies are in eV.}
\footnotetext{$^a$For the negative ions, the orbital basis was aug-cc-pVTZ.}
\label{table:t}
\end{table}

\color{black}
In concluding this section, we first recall the reasons why our CDFA method was implemented with the positivity constraint \eqref{pos}. This constraint is intuitive if one 
considers each electron to experience a repulsive electronic density from the other $N-1$ electrons, but it also serves two computational purposes: (i) to avoid shifting negative 
screening density to infinity as the size of the orbital and auxiliary basis sets increase and (ii) as a regularization technique to avoid pathological behaviour of the OEP solution. 
As we have seen from the good convergence of the screening densities and potentials, the latter reason is no longer necessary with the introduction of the complement terms in the 
OEP equation \eqref{oep_l}.

Regarding the first reason (i), we note that for the moderately large auxiliary basis sets we tested, it is safe to carry out constrained minimisations of the DFA total energy under the constraint of the screening charge only, $Q_\textrm{scr} = N-1$, except for systems with few electrons; for these systems the omission of the positivity constraint manifests itself in shifting negative screening density away from the origin. As such, the benefits of removing the positivity condition which is a computational bottleneck usually outweigh the disadvantages. For the benefit of readers less familiar with OEP calculations, we outline the full simplified procedure for solving the CDFA equations in Appendix~\ref{AppA}.

\color{black}
In the next section, we shall argue that the screening charge constraint endows the xc potential of local and semi-local DFAs with exotic qualities,
such a finite derivative discontinuity $\Delta_\textrm{xc}$. Although crucial for the accurate prediction of band gaps, $\Delta_\textrm{xc}$ is notoriously 
hard to capture in approximations. Advanced approximations have been proposed which capture this discontinuous behaviour, e.g., 
Refs.~\onlinecite{andrade,kraisler2014fundamental,PhysRevA.98.022513,senjean2020n,guandalini}, 
however, further development is required for these methods to yield reliable band gaps for all materials.

\section{Derivative discontinuity of the CDFA xc potential}

The discontinuity of the xc potential is defined by
\begin{equation} \label{Dxc_def}
\Delta_\textrm{xc} = \lim_{\omega \rightarrow 0^+ } \Delta_\textrm{xc}^{\omega} ( \rr ) , \ \  {\rm with} \ 
\Delta_\textrm{xc}^{\omega} ( \rr ) = v_\textrm{xc}^{N+\omega} ( \rr ) -  v_\textrm{xc}^{N-\omega} ( \rr )
\end{equation}
where $v_\textrm{xc}^{N\pm\omega} ( \rr )$ is the xc potential of an ensemble with $N\pm\omega$ electrons. 

The ensemble KS densities with $N\pm \omega$ electrons are given by,
\begin{eqnarray}
\rho^{N-\omega}_{v_\textrm{en}} ( \rr ) & = & \omega \rho^{N-1}_{v_\textrm{en}} ( \rr ) + (1-\omega) \rho^{N}_{v_\textrm{en}} ( \rr ) , \label{eq19} \\
\rho^{N+\omega}_{v_\textrm{en}} ( \rr ) & = & (1-\omega) \rho^N_{v_\textrm{en}} ( \rr ) + \omega \rho^{N+1}_{v_\textrm{en}} ( \rr ) , \label{eq20}
\end{eqnarray}
where $\rho^{M}_{v_\textrm{en}} ( \rr )$, $M=N-1,N,N+1$, is the ground state density of the $M$-electron KS system 
in the external potential $v_\textrm{en} ( \rr )$. We shall use the CLDA KS equation \eqref{ks_eq}, with constraints (\ref{norm},\ref{pos}). 

We seek  the derivative discontinuity $\Delta_\textrm{xc}$ of the  CLDA xc potential from \eqref{Dxc_def} for reference.
In order to obtain $\Delta_\textrm{xc}^{\omega} ( \rr )$ and then $\Delta_\textrm{xc}$, one must first find the ensemble KS xc potentials with densities 
$\rho^{N \pm \omega}_{v_\textrm{en}} ( \rr )$ and subtract them. 
Work is in progress in our group to obtain directly these ensemble KS xc potentials.
Here, we use the method of Refs.~\onlinecite{Hodgson_ensembles} and \onlinecite{kraisler2021kohnsham} to obtain the ensemble KS xc potential by constructing the ensemble density $\rho_{v_\textrm{en}}^{N\pm \omega}$ 
from separate KS calculations for $N$, and $N \pm 1$ particles and then inverting $\rho_{v_\textrm{en}}^{N\pm \omega} ( \rr ) $ to obtain
$v_\textrm{xc} ^{N\pm \omega} ( \rr )$.

Let us follow this construction in detail. 
The two KS ground state densities that build the ensemble density $\rho_{v_\textrm{en}}^{N + \omega} ( \rr ) $ can be written:
\begin{eqnarray}
\rho^N_{v_\textrm{en}} ( \rr ) & = & \sum_{i = 1}^N | \phi_i [\rho^N] ( \rr ) |^2 \label{eq21}\\
\rho^{N+1}_{v_\textrm{en}} ( \rr ) & = & \sum_{i = 1}^{N+1} | \phi_i [\rho^{N+1}] ( \rr ) |^2 \label{eq22} 
\end{eqnarray}
The notation makes explicit that $\{ \phi_i [\rho^M] ( \rr ) \}$ are the KS orbitals of the $M$-electron system with density $\rho^M$.

In terms of the ensemble KS orbitals $\{ \phi_i [\rho^{N+\omega}] (\bm r) \}$ the ensemble density is given by
\begin{equation} \label{rho_ens2}
\rho^{N+\omega}_{v_\textrm{en}} ( \rr )  =  \sum_{i = 1}^N
| \phi_i [\rho^{N+\omega}] ( \rr ) |^2 
+ \omega \, | \phi_{N+1} [\rho^{N+\omega}] ( \rr ) |^2 .
\end{equation}
In addition, from Eqs.~\ref{eq19}-\ref{eq22}, it is also equal to 
\begin{equation} \label{rho_ens}
\rho^{N+\omega}_{v_\textrm{en}} ( \rr )  =  \sum_{i = 1}^N
\bigg[ (1-\omega) | \phi_i [\rho^N] ( \rr ) |^2 + \omega | \phi_i [\rho^{N+1}] ( \rr ) |^2  \bigg]
+ \omega \, | \phi_{N+1} [\rho^{N+1}] ( \rr ) |^2 .
\end{equation}
In general, the ensemble KS orbitals, $\{ \phi_i [\rho^{N + \omega} ] ( \rr ) \}$ in \eqref{rho_ens2}, 
will be linear combinations of the two sets of KS orbitals in \eqref{rho_ens}.
However, in the asymptotic region the picture is very simple.
For any $\omega > 0$, the density $| \phi_{N+1} [\rho^{N+1}] ( \rr ) |^2$ 
of the $N+1$ orbital will be the dominant term as every other term of Eq.~(\ref{rho_ens}) in the ensemble density will have died out. 
Hence the tail of the ($N+1$)-th ensemble-KS orbital of Eq.~(\ref{rho_ens2}), $\phi_{N+1} [\rho^{N+\omega}] ( \rr )$, 
will be equal, within a phase, to the tail of $\phi_{N+1} [\rho^{N+1}] ( \rr )$. 
However, $\phi_{N+1} [\rho^{N+1}] ( \rr )$ is a KS orbital of the $N+1$ electron system and in the asymptotic region it feels the net 
Coulomb repulsion of a screening charge of $N$ electrons. 
Consequently, $\phi_{N+1} [\rho^{N+\omega}] ( \rr )$, in the asymptotic region, 
must feel the Coulomb repulsion of an equal amount of screening charge. Since the ensemble-KS orbitals lie in a common KS potential,
the screening charge of the ensemble-screening-density will be $Q_\textrm{scr}^{N+\omega} = N$, for any finite $\omega > 0$.

We conclude that when the number of electrons increases past an integer value, the value of the screening charge 
$Q_\textrm{scr}^{N+\omega}$ increases stepwise, 
\begin{equation} \label{qscr_ens}
Q_\textrm{scr}^{M+\omega} = M , \ {\rm with} \ M = N , N \pm 1 , \ldots \ {\rm and} \ 0 < \omega \le 1 .
\end{equation}
In the limit $\omega \rightarrow 0^+$, we have: 
\begin{equation}
Q_\textrm{scr}^{N} = N-1 , \ \ Q_\textrm{scr}^{N^+} = N ,
\end{equation}
where $ Q_\textrm{scr}^{N^+} = \lim_{\omega \rightarrow 0^+} Q_\textrm{scr}^{N+\omega} $.

This stepwise increase of screening charge obviously causes a discontinuous jump in the constrained xc potential $v_\textrm{xc}^{N+\omega} ( \rr ) $. 
In the limit $\omega \rightarrow 0^+$, the jump of the xc potential is $v_\textrm{xc}^{N^+} ( \rr ) - v_\textrm{xc}^{N} ( \rr ) $, 
where $v_\textrm{xc}^{N^+} ( \rr ) = \lim_{\omega \rightarrow 0^+} v_\textrm{scr}^{N+\omega} ( \rr ) $.
From \eqref{Dxc_def} the jump of the xc potential due to the stepwise increase in the screening charge gives the 
derivative discontinuity in the CDFA method,
\begin{equation} \label{dxc0}
\Delta_\textrm{xc}^\textrm{CDFA} (\rr) = v_\textrm{xc}^{N^+} ( \rr ) - v_\textrm{xc}^{N} ( \rr ) .
\end{equation}
We note that Eq.~\ref{dxc0} does not require an ensemble calculation, but only the evaluation of the $N$-electron CDFA xc potential for two values of the screening charge and hence could be employed in practical calculations at a moderate computational cost.  

In the last part of the paper, we shall compare $\Delta_\textrm{xc} $ from the constrained minimisation method \eqref{dxc0} with the 
result for $\Delta_\textrm{xc} $ from \eqref{Dxc_def}.
We shall calculate the differences
\begin{equation} \label{dxcw}
 \Delta_\textrm{xc} ^ \omega ( \rr ) \simeq  v_\textrm{xc}^{N+\omega} ( \rr ) -  v_\textrm{xc}^{N} ( \rr )
\end{equation}
in CLDA for various values of $\omega$ and investigate the limit of small $\omega$.

\color{black}
Before we continue, we note that in the simple model we have constructed to predict the derivative discontinuity, 
using the inversion of the ensemble density \eqref{Dxc_def} and with the CDFA method \eqref{dxc0}, 
we have restricted the freedom of the Hxc potentials, by the ansatz in \eqref{v_rho}; the restriction is that $v_{\rm xc}^{N+\omega} ( \infty ) = 0$.
Consequently, the derivative discontinuities we obtain with Eq.~\ref{Dxc_def} and Eq.~\ref{dxc0} cannot be perfect constant functions 
but have to vanish at $r \rightarrow \infty$. 
We aim to investigate whether the resulting approximate derivative discontinuity, $\Delta_{\rm xc} (\rr)$, as a function of $\rr$ remains 
flat and almost equal to a constant over the region of the atom or the molecule. Finally, we want 
to obtain the converged value of the constant in the limit of an infinite basis set. 

\color{black}
In order to proceed and construct the ensemble density $\rho^{N+\omega}_{v_\textrm{en}} ( \rr ) $, 
we need the densities from two KS calculations for $N$ and $N+1$ particles 
allowing us to then find the corresponding ensemble xc potential against which $\Delta_\textrm{xc}^\textrm{CDFA}$ can be compared.
We use our CLDA method to obtain the densities  $\rho^{N}_{v_\textrm{en}} ( \rr )$ and $\rho^{N+1}_{v_\textrm{en}} ( \rr )$, in order to control
the screening densities of the constituent xc potentials.
One of the integers $N$, $N+1$ is an odd number, corresponding to an open shell system. 
The LDA exchange energy for open shells contains an error (``ghost-exchange error'' \cite{ilda}) in modelling exchange with half the electrons spin-up 
and half spin-down. In a forthcoming publication \cite{ilda}, we propose how to correct this error, still within LDA (not local spin density approximation). 
Hence, in the KS calculation for an odd number of electrons (either for $N$ or for $N+1$), 
we employ our method to correct for the ghost-exchange error, in order to improve the accuracy of the resulting CLDA xc potential and density. 
Details will be published in Ref.~\onlinecite{ilda}.

Once we construct the ensemble density, we invert it to obtain the ensemble KS potential, $v_\textrm{xc}^{N+\omega} ( \rr ) $. For the inversion, we 
apply the method in Refs.~\onlinecite{invert,lfx}. 
The inversion method \cite{invert} requires the a priori selection of a value for the screening charge of the xc potential. 
According to \eqref{qscr_ens}, for $v_\textrm{xc}^{N+\omega} ( \rr ) $ we set $Q_\textrm{scr}^{N+\omega} = N$.

\begin{figure}[h!]
  \centering
  \subfloat[$v_\textrm{xc}^{N+\omega} ( r ) $]
  {\includegraphics[trim=0 0 0 0, clip, width=0.45\textwidth]{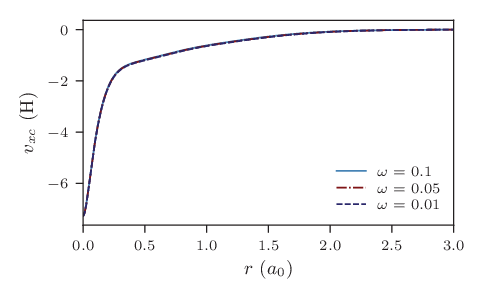}\label{fig:9a}}
  \subfloat[$r^2 \, \rho_\textrm{scr}^{N+\omega} (r)$]
  {\includegraphics[trim=0 0 0 0, clip, width=0.45\textwidth]{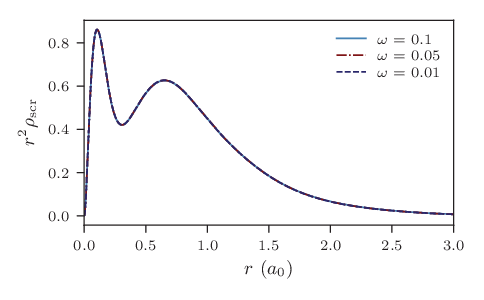}\label{fig:9b}}
   \caption{Ne atom, ensemble xc potentials and screening densities for various values of $\omega$. 
The orbital and auxiliary basis sets are un-contracted cc-pVTZ.}
   \label{fig:Clr2}
\end{figure}

\begin{figure}[h!]
  \centering
  \subfloat[xc potentials $v_\textrm{xc}^{N+\omega} (r)$ and $v_\textrm{xc}^N(r)$, $v_\textrm{xc}^{N^+}(r)$.]
  {\includegraphics[trim=0 0 0 0, clip, width=0.5\textwidth]{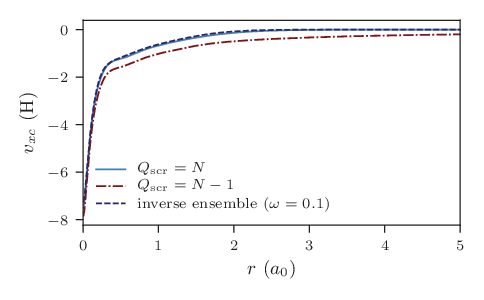}\label{fig:10a}}
  \subfloat[$\Delta_\textrm{xc}^\omega ( r ) $ for various $\omega$. 
Blue line is $\Delta_\textrm{xc}^\textrm{CLDA} (r)$.]
  {\includegraphics[trim=0 0 0 0, clip, width=0.5\textwidth]{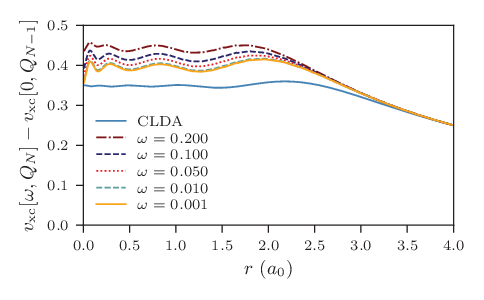}\label{fig:10b}}
   \caption{Ne atom, xc potentials and differences of xc potentials.
The orbital and auxiliary basis sets are un-contracted cc-pVTz.}
   \label{fig:Clr3}
\end{figure}

In Figs.~\ref{fig:9a}, \ref{fig:9b}  the ensemble xc potentials, $v_\textrm{xc}^{N+\omega} ( r ) $ and screening densities
are shown, for various values of $\omega$, obtained by inverting the ensemble densities \eqref{rho_ens}. 
The screening charge for the ensemble densities is set $Q_\textrm{scr}^{N+\omega} = N$. 
The xc potentials and screening densities are very close, as expected, 
which is an indication of the quality of convergence and the inversion method.

In Fig.~\ref{fig:10a}, the ensemble xc potential, $v_\textrm{xc}^{N+\omega} ( r )$,  for $\omega=0.1$ (with $Q_\textrm{scr}^{N+\omega} = N$)
is shown together with $v_\textrm{xc}^{N} ( r ) $ and $v_\textrm{xc}^{N^+} ( r )$, which have screening charges 
$Q_\textrm{scr}^{N} = N-1$ and $Q_\textrm{scr}^{N^+} = N$. In Fig.~\ref{fig:10b}, the $\omega$-dependent \eqref{dxcw} derivative discontinuity, 
$\Delta_\textrm{xc}^{\omega} ( r ) =  v_\textrm{xc}^{N+\omega} ( r ) - v_\textrm{xc}^N ( r ) $, is shown for various values of $\omega$. 
In the limit of small $\omega$, $\Delta_\textrm{xc}^{\omega} ( r )$ yields the derivative discontinuity using ensembles, 
$\Delta_\textrm{xc}^{\omega \rightarrow 0} ( r ) = 
 \Delta_\textrm{xc} (r)$. 

\color{black}
The inversion method has some numerical instabilities which are exaggerated when the difference of two potentials is taken. This explains why 
$\Delta_\textrm{xc} ( r ) $ is not flat for small $r$. The distance $r$ after which $\Delta_\textrm{xc} ( r ) $ tends to zero depends on the basis set. 
However, we do not propose this method as a means of computing the derivative discontinuity in practice, but rather to compare with the results of the CDFA method.

\color{black}
The blue line in Fig.~\ref{fig:10b} shows the CLDA prediction for the derivative discontinuity, $\Delta_\textrm{xc}^\textrm{CLDA} (r)$, 
without an ensemble calculation. 
$\Delta_\textrm{xc}^\textrm{CLDA} (r)$ remains almost a constant up to a distance of about 2.5 $a_0$, beyond which it tends to zero. 
%
The differences $\Delta_\textrm{xc}^\omega ( r ) $ for decreasing $\omega$ approach the line of $\Delta_\textrm{xc}^\textrm{CLDA}$
both in height and in the spatial extent over which $\Delta_\textrm{xc}^\textrm{CLDA}$ and $\Delta_\textrm{xc}^\omega$ stay almost constant. 
\color{black}
The value of the constant can be obtained by inspection from Fig.~\ref{fig:10b} to be approximately 0.35 hartrees, or about 9.5eV.  
We can obtain the constant more accurately from the shift of the occupied single-particle energy levels between the two xc potentials 
$v_{\rm xc}^N (\rr)$ and $v_{\rm xc}^{N^+} ( \rr )$. For the un-contracted cc-pVTz basis used for the results in Fig.~\ref{fig:10b} we find 
$\Delta_{\rm xc}^{\rm CLDA} = 9.48$eV. See Table~\ref{table:t2}.

We conclude this section by investigating the dependence of basis set size on (a) the height of $\Delta_{\rm xc}^{\rm CDFA} ( r )$, 
in the region where it is stays almost flat and (b) the spatial extent of the region over which $\Delta_{\rm xc}^{\rm CDFA} ( r )$ remains flat.

We calculated the xc derivative discontinuity with our model $\Delta_{\rm xc}^{\rm CDFA} ( r )$ \eqref{dxc0} using as 
orbital and auxiliary basis sets the un-contracted cc-pVXz sets, with X=D,T,Q,5. 
The last row of Table~\ref{table:t2} shows the value of the derivative discontinuity, $\Delta_{\rm xc}^{\rm CDFA} $, for each basis set.  

Each column in Table~\ref{table:t2} shows the eigenvalues of the occupied orbitals in the Ne atom, with the two constrained 
xc potentials $v_{\rm xc}^N (\rr)$ 
and $v_{\rm xc}^{N^+} ( \rr )$, for a specific choice of orbital and auxiliary basis sets.
Each column also shows the shift of each eigenvalue $\Delta_i = \epsilon_i^{N^+} - \epsilon_i^N$. The average value 
of these shifts, gives $\Delta_{\rm xc}^{\rm CLDA}$ in the specific basis.  
\begin{table}[h!]
\begin{tabular}{ | c  | c |  c | c | c |}
 \hline
& cc-pVDz  & cc-pVTz & cc-pVQz & cc-pV5z \\
\hline
$\epsilon_{\rm 1s}^N $ &-830.10& -829.96   &  -830.60  & -829.63    \\
$\epsilon_{\rm 1s}^{N^+} $ &-817.86 & -820.48    &  -822.86  & -823.02    \\
$\Delta_{\rm 1s} $  & 12.24 & 9.48 & 7.74&6.61 \\
\hline
$\epsilon_{\rm 2s}^N $ & -40.68 & -41.14.  &  -41.91  &  -40.97   \\
$\epsilon_{\rm 2s}^{N^+} $ & -28.45 & -31.66  &  -34.17  &  -34.35    \\
$\Delta_{\rm 2s} $ &12.23& 9.48 & 7.74 & 6.62 \\
\hline
$\epsilon_{\rm 2p}^N $ & -18.07 & -18.65  &  -19.44  &  -18.52   \\
$\epsilon_{\rm 2p}^{N^+} $ & -5.86 & -9.17  &  -11.71 & -11.9      \\
$\Delta_{\rm 2p} $ &12.21& 9.48 & 7.73 & 6.62\\
\hline
$\Delta_{\rm xc}^{\rm CLDA} $ & 12.22 & 9.48 & 7.73 & 6.62 \\
\hline
\end{tabular}
\caption{Ne atom. The bound eigenvalues $\epsilon_i^N$, $\epsilon_i^{N^+}$, and their difference $\Delta_i = \epsilon_i^{N^+} - \epsilon_i^{N}$, 
of the CLDA xc potentials $v_{\rm xc}^{N} $, $v_{\rm xc}^{N^+} $, for the orbitals 
$i = $ 1s, 2s, 2p. The orbital and auxiliary basis set is un-contracted cc-pVXz, X=D,T,Q,5. 
The average difference $\Delta_i$ per basis set gives $\Delta_{\rm xc}^{\rm CLDA}$. All energies are in eV.
}
\label{table:t2}
\end{table}
Using the un-contracted cc-pVDz orbital and auxiliary basis, the shifts of the orbital eigenvalues are almost the same within 0.03eV. 
In the un-contracted cc-pVTz orbital and auxiliary basis, the differences in the shifts in each energy level are
smaller than 0.01eV. In the two larger basis sets, the differences between the almost constant shifts for each energy level are 
within 0.01eV. These results are consistent with a near perfectly constant $\Delta_{\rm xc}^{\rm CLDA}$ over the whole spatial region 
where the electronic density of the Ne atom is appreciable.  

For the un-contracted cc-pVTz basis, we performed another check to confirm that the shift between the two xc potentials 
$v_{\rm xc}^{N^+} ( \rr) $, $v_{\rm xc}^N ( \rr) $ is almost a constant over a large region of space.
We evaluated the overlaps of the occupied orbitals in the two potentials, $\langle \phi_i^{N^+} | \phi_i^N \rangle $, $i=$1s, 2s, 2p (triply degenerate). 
We found that the numerical values of all overlaps were indeed very close to one, with the overlap in the worst case differing from one by about $\sim 10^{-7}$. 

\begin{figure}[h!]
  \centering
  \subfloat[Function $\Delta_{\rm xc}^{\rm CLDA} ( r )$ for various basis sets.]
  {\includegraphics[trim=0 0 0 0, clip, width=0.5\textwidth]{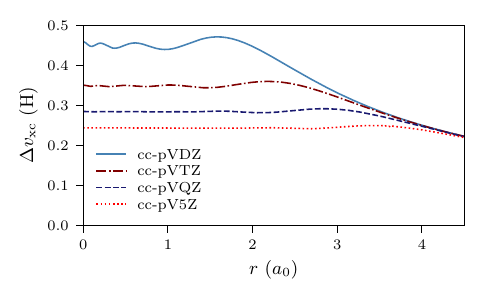}\label{fig:11a}}
  \subfloat[Extraploation of $\Delta_{\rm xc}^{\rm CLDA} $ vs basis set size $n_{\rm bas}$.]
  {\includegraphics[trim=0 0 0 0, clip, width=0.5\textwidth]{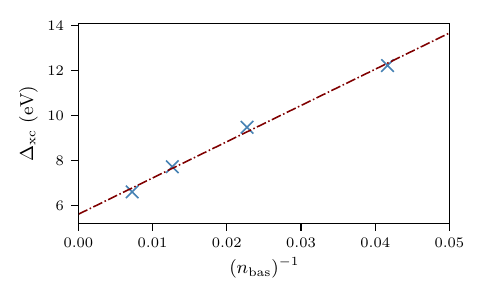}\label{fig:11b}}
   \caption{Ne atom. Left: The xc derivative discontinuity (height of plateau in $\Delta_{\rm xc}^{\rm CLDA} ( r ) $) 
   decreases while the extent of plateau increases with increasing basis set size. Right: 
   $\Delta_{\rm xc}^{\rm CLDA}$ behaves linearly with inverse basis set size ($n_{\rm bas}$ is the number of basis set elements).
   The extrapolation line intersects the vertical axis (infinite basis set limit) at $\Delta_{\rm xc}^{\rm CLDA} = 5.6$eV.}
   \label{fig:Clr3}
\end{figure}

In Fig.~\ref{fig:11a} we show the derivative discontinuity $\Delta_{\rm xc}^{\rm CLDA} ( r ) $ as a function of $r$ (Eq.~\ref{dxc0}) 
for orbital and auxiliary basis sets un-contracted cc-pVXz, X=D,T,Q,5. These functions have a plateau at the origin where the atom lies. The extent of the 
plateau increases with basis set size and the height decreases and seems to converge.
To establish that the discontinuity $\Delta_{\rm xc}^{\rm CLDA} $ (height of the plateau) indeed converges and does not vanish in the limit of infinite basis set, we plot 
$\Delta_{\rm xc}^{\rm CLDA}$ against the inverse of the number of basis set elements, $n_{\rm bas}$. The behaviour is fitted well by a straight line with equation
$\Delta_{\rm xc}^{\rm CLDA} ( n_{\rm bas} ) = 5.6 + 160\times (n_{\rm bas})^{-1}$. The extrapolation gives a nonzero derivative discontinuity of 5.6eV for the infinite basis limit.

\color{black}
\section{Conclusions}

A common theme of popular local and semi-local density functional approximations is the imbalance of accuracy between 
energy density-functionals, which can be quite accurate, and the corresponding effective KS potentials, with inferior accuracy 
\cite{adventures_bartlett,inconsistent,kieron_density_error}. 
We have approached this problem from several directions \cite{gidopoulos_ks,lfx,callow_gidopoulos}. 
In this paper, we review and expand our work on imposing physical constraints during the energy minimisation in order to yield more a accurate corresponding xc potential \cite{clda,clda_review,cdfa}. 
Specifically, we investigate the relaxation of a constraint that is computationally expensive and find that its omission leads to well behaved results,  
except for very small systems with only a few electrons. 
The constraints we impose raise the total energy minimally \cite{clda,cdfa} but have a dramatic impact on the quality of the effective KS potential, 
gifting it with the correct asymptotic behaviour and enabling it to exhibit important non-analytic behaviour (derivative discontinuity) 
shared by the exact KS potential but elusive from the lower rungs of Jacob's ladder of DFAs where semi-local DFAs reside.

\color{black}
\appendix 
\section{} \label{AppA}

Below we summarize the full computational procedure for the constrained DFA method described in Sec.~\ref{sec:2a}.
\begin{enumerate}
    \item Make an initial guess for the KS orbitals and the screening density, which is expanded in the auxiliary basis set,
    \begin{equation}
        \rho_\textrm{scr}(\rr)=\sum_k \rho^\textrm{s}_k \theta_k(\rr).
    \end{equation}
    \item Construct the matrices
    \begin{align}
        A_{kl}&= \mel*{\tilde{\theta_k}}{\chi^0}{\tilde{\theta_l}} + \lambda  \mel*{\tilde{\theta_k}}{\bar{\chi}}{\tilde{\theta_l}},\\
        b_k&= \ip*{\tilde{\theta_k}}{b^0} + \lambda  \ip*{\tilde{\theta_k}}{\bar{b}},\ \textrm{where}\ \tilde{\theta}_k (\rr) = \int\dd{\rr'} \frac{\theta_k(\rr')}{|\rr-\rr'|}.
    \end{align}
    The vector $b_k$ contains information about the functional being used (such as LDA), as seen in Eq.~\eqref{eq:b0}.
    \item Solve the OEP matrix equation,
    \begin{equation}
        \sum_l A_{kl} \rho^\textrm{s}_l = b_k,
    \end{equation}
    to obtain the updated coefficients $\rho^\textrm{s}_l$, under the constraint that $Q_\textrm{scr}=N-1$,
    \begin{equation}
        \sum_k \rho^\textrm{s}_k X_k = Q_\textrm{scr},\ X_k = \int\dd{\rr} \theta_k (\rr).
    \end{equation}
    This is equivalent to solving the equations
    \begin{align}
        \rho^\textrm{s}_k &= \sum_l (A)^{-1}_{kl} (b_l + \alpha X_l), \\
        \alpha &= \frac{Q_\textrm{scr} - \sum_{kl}X_k A_{kl} b_l}{\sum_k X_k^2}.
    \end{align}
    \item With the new Hxc-potential constructed via the screening density from the previous step, diagonalize the KS Fock matrix to update the KS orbitals.
    \item Repeat steps 2-4 until the energy and density matrix are converged.
\end{enumerate}
  
\begin{acknowledgements}
N.I.G. and T.P. acknowledge financial support by The Leverhulme Trust, through a Research Project Grant with number RPG-2016-005.\\
N.I.G. thanks Prof. Rod Bartlett for helpful discussions during his visit at Durham University in early 2019 and acknowledge 
the Institute of Advanced Study at Durham University for hosting this visit. N.I.G. and T.J.C. thank Prof. E.K.U. Gross for helpful discussions. 
M.J.P.H. gratefully acknowledges support from Prof. E.K.U. Gross.
\end{acknowledgements}

\color{black}
  

\begin{thebibliography}{50}
\expandafter\ifx\csname natexlab\endcsname\relax\def\natexlab#1{#1}\fi
\expandafter\ifx\csname bibnamefont\endcsname\relax
  \def\bibnamefont#1{#1}\fi
\expandafter\ifx\csname bibfnamefont\endcsname\relax
  \def\bibfnamefont#1{#1}\fi
\expandafter\ifx\csname citenamefont\endcsname\relax
  \def\citenamefont#1{#1}\fi
\expandafter\ifx\csname url\endcsname\relax
  \def\url#1{\texttt{#1}}\fi
\expandafter\ifx\csname urlprefix\endcsname\relax\def\urlprefix{URL }\fi
\providecommand{\bibinfo}[2]{#2}
\providecommand{\eprint}[2][]{\url{#2}}

\bibitem[{\citenamefont{Bartlett}(2019)}]{adventures_bartlett}
\bibinfo{author}{\bibfnamefont{R.~J.} \bibnamefont{Bartlett}},
  \bibinfo{journal}{The Journal of Chemical Physics}
  \textbf{\bibinfo{volume}{151}}, \bibinfo{pages}{160901}
  (\bibinfo{year}{2019}), \urlprefix\url{https://doi.org/10.1063/1.5116338}.

\bibitem[{\citenamefont{Wasserman et~al.}(2017)\citenamefont{Wasserman,
  Nafziger, Jiang, Kim, Sim, and Burke}}]{inconsistent}
\bibinfo{author}{\bibfnamefont{A.}~\bibnamefont{Wasserman}},
  \bibinfo{author}{\bibfnamefont{J.}~\bibnamefont{Nafziger}},
  \bibinfo{author}{\bibfnamefont{K.}~\bibnamefont{Jiang}},
  \bibinfo{author}{\bibfnamefont{M.-C.} \bibnamefont{Kim}},
  \bibinfo{author}{\bibfnamefont{E.}~\bibnamefont{Sim}}, \bibnamefont{and}
  \bibinfo{author}{\bibfnamefont{K.}~\bibnamefont{Burke}},
  \bibinfo{journal}{Annual Review of Physical Chemistry}
  \textbf{\bibinfo{volume}{68}}, \bibinfo{pages}{555} (\bibinfo{year}{2017}),
  \bibinfo{note}{pMID: 28463652},
  \urlprefix\url{https://doi.org/10.1146/annurev-physchem-052516-044957}.

\bibitem[{\citenamefont{Sim et~al.}(2018)\citenamefont{Sim, Song, and
  Burke}}]{kieron_density_error}
\bibinfo{author}{\bibfnamefont{E.}~\bibnamefont{Sim}},
  \bibinfo{author}{\bibfnamefont{S.}~\bibnamefont{Song}}, \bibnamefont{and}
  \bibinfo{author}{\bibfnamefont{K.}~\bibnamefont{Burke}},
  \bibinfo{journal}{The journal of physical chemistry letters}
  \textbf{\bibinfo{volume}{9}}, \bibinfo{pages}{6385} (\bibinfo{year}{2018}).

\bibitem[{\citenamefont{Gidopoulos}(2011)}]{gidopoulos_ks}
\bibinfo{author}{\bibfnamefont{N.~I.} \bibnamefont{Gidopoulos}},
  \bibinfo{journal}{Phys. Rev. A} \textbf{\bibinfo{volume}{83}},
  \bibinfo{pages}{040502} (\bibinfo{year}{2011}),
  \urlprefix\url{https://link.aps.org/doi/10.1103/PhysRevA.83.040502}.

\bibitem[{\citenamefont{Hollins et~al.}(2016)\citenamefont{Hollins, Clark,
  Refson, and Gidopoulos}}]{lfx}
\bibinfo{author}{\bibfnamefont{T.}~\bibnamefont{Hollins}},
  \bibinfo{author}{\bibfnamefont{S.}~\bibnamefont{Clark}},
  \bibinfo{author}{\bibfnamefont{K.}~\bibnamefont{Refson}}, \bibnamefont{and}
  \bibinfo{author}{\bibfnamefont{N.}~\bibnamefont{Gidopoulos}},
  \bibinfo{journal}{Journal of Physics: Condensed Matter}
  \textbf{\bibinfo{volume}{29}}, \bibinfo{pages}{04LT01}
  (\bibinfo{year}{2016}),
  \urlprefix\url{http://stacks.iop.org/0953-8984/29/i=4/a=04LT01}.

\bibitem[{\citenamefont{Callow and Gidopoulos}(2018)}]{callow_gidopoulos}
\bibinfo{author}{\bibfnamefont{T.~J.} \bibnamefont{Callow}} \bibnamefont{and}
  \bibinfo{author}{\bibfnamefont{N.~I.} \bibnamefont{Gidopoulos}},
  \bibinfo{journal}{The European Physical Journal B}
  \textbf{\bibinfo{volume}{91}}, \bibinfo{pages}{209} (\bibinfo{year}{2018}).

\bibitem[{\citenamefont{Irons et~al.}(2017)\citenamefont{Irons, Furness, Ryley,
  Zemen, Helgaker, and Teale}}]{helgaker_teale}
\bibinfo{author}{\bibfnamefont{T.~J.} \bibnamefont{Irons}},
  \bibinfo{author}{\bibfnamefont{J.~W.} \bibnamefont{Furness}},
  \bibinfo{author}{\bibfnamefont{M.~S.} \bibnamefont{Ryley}},
  \bibinfo{author}{\bibfnamefont{J.}~\bibnamefont{Zemen}},
  \bibinfo{author}{\bibfnamefont{T.}~\bibnamefont{Helgaker}}, \bibnamefont{and}
  \bibinfo{author}{\bibfnamefont{A.~M.} \bibnamefont{Teale}},
  \bibinfo{journal}{The Journal of chemical physics}
  \textbf{\bibinfo{volume}{147}}, \bibinfo{pages}{134107}
  (\bibinfo{year}{2017}).

\bibitem[{\citenamefont{Gidopoulos and Lathiotakis}(2012{\natexlab{a}})}]{clda}
\bibinfo{author}{\bibfnamefont{N.}~\bibnamefont{Gidopoulos}} \bibnamefont{and}
  \bibinfo{author}{\bibfnamefont{N.}~\bibnamefont{Lathiotakis}},
  \bibinfo{journal}{J. Chem. Phys.} \textbf{\bibinfo{volume}{136}},
  \bibinfo{pages}{224109} (\bibinfo{year}{2012}{\natexlab{a}}),
  \urlprefix\url{https://doi.org/10.1063/1.4728156}.

\bibitem[{\citenamefont{Gidopoulos and Lathiotakis}(2015)}]{clda_review}
\bibinfo{author}{\bibfnamefont{N.}~\bibnamefont{Gidopoulos}} \bibnamefont{and}
  \bibinfo{author}{\bibfnamefont{N.~N.} \bibnamefont{Lathiotakis}}, in
  \emph{\bibinfo{booktitle}{Advances In Atomic, Molecular, and Optical
  Physics}}, edited by
  \bibinfo{editor}{\bibfnamefont{E.}~\bibnamefont{Arimondo}},
  \bibinfo{editor}{\bibfnamefont{C.~C.} \bibnamefont{Lin}}, \bibnamefont{and}
  \bibinfo{editor}{\bibfnamefont{S.~F.} \bibnamefont{Yelin}}
  (\bibinfo{publisher}{Academic Press}, \bibinfo{year}{2015}),
  vol.~\bibinfo{volume}{64}, pp. \bibinfo{pages}{129 -- 142},
  \urlprefix\url{http://www.sciencedirect.com/science/article/pii/S1049250X15000063}.

\bibitem[{\citenamefont{Pitts et~al.}(2018)\citenamefont{Pitts, Gidopoulos, and
  Lathiotakis}}]{cdfa}
\bibinfo{author}{\bibfnamefont{T.}~\bibnamefont{Pitts}},
  \bibinfo{author}{\bibfnamefont{N.~I.} \bibnamefont{Gidopoulos}},
  \bibnamefont{and} \bibinfo{author}{\bibfnamefont{N.~N.}
  \bibnamefont{Lathiotakis}}, \bibinfo{journal}{Eur. Phys. J. B}
  \textbf{\bibinfo{volume}{91}}, \bibinfo{pages}{130} (\bibinfo{year}{2018}),
  \urlprefix\url{https://doi.org/10.1140/epjb/e2018-90123-8}.

\bibitem[{\citenamefont{Perdew and Schmidt}(2001)}]{jacob_ladder}
\bibinfo{author}{\bibfnamefont{J.}~\bibnamefont{Perdew}} \bibnamefont{and}
  \bibinfo{author}{\bibfnamefont{K.}~\bibnamefont{Schmidt}},
  \emph{\bibinfo{title}{Density Functional Theory and its Applications to
  Materials}} (\bibinfo{publisher}{American Institute of Physics},
  \bibinfo{address}{Melville, New York}, \bibinfo{year}{2001}), vol.
  \bibinfo{volume}{577} of \emph{\bibinfo{series}{API Conference Proceedings}},
  pp. \bibinfo{pages}{1--20}.

\bibitem[{\citenamefont{Sharp and Horton}(1953)}]{sharp}
\bibinfo{author}{\bibfnamefont{R.~T.} \bibnamefont{Sharp}} \bibnamefont{and}
  \bibinfo{author}{\bibfnamefont{G.~K.} \bibnamefont{Horton}},
  \bibinfo{journal}{Phys. Rev.} \textbf{\bibinfo{volume}{90}},
  \bibinfo{pages}{317} (\bibinfo{year}{1953}),
  \urlprefix\url{https://link.aps.org/doi/10.1103/PhysRev.90.317}.

\bibitem[{\citenamefont{Talman and Shadwick}(1976)}]{talman}
\bibinfo{author}{\bibfnamefont{J.~D.} \bibnamefont{Talman}} \bibnamefont{and}
  \bibinfo{author}{\bibfnamefont{W.~F.} \bibnamefont{Shadwick}},
  \bibinfo{journal}{Phys. Rev. A} \textbf{\bibinfo{volume}{14}},
  \bibinfo{pages}{36} (\bibinfo{year}{1976}),
  \urlprefix\url{https://link.aps.org/doi/10.1103/PhysRevA.14.36}.

\bibitem[{\citenamefont{G\"orling}(1999)}]{gorling_poisson}
\bibinfo{author}{\bibfnamefont{A.}~\bibnamefont{G\"orling}},
  \bibinfo{journal}{Phys. Rev. Lett.} \textbf{\bibinfo{volume}{83}},
  \bibinfo{pages}{5459} (\bibinfo{year}{1999}),
  \urlprefix\url{https://link.aps.org/doi/10.1103/PhysRevLett.83.5459}.

\bibitem[{\citenamefont{Liu et~al.}(1999)\citenamefont{Liu, Ayers, and
  Parr}}]{liu_ayers_parr}
\bibinfo{author}{\bibfnamefont{S.}~\bibnamefont{Liu}},
  \bibinfo{author}{\bibfnamefont{P.~W.} \bibnamefont{Ayers}}, \bibnamefont{and}
  \bibinfo{author}{\bibfnamefont{R.~G.} \bibnamefont{Parr}},
  \bibinfo{journal}{The Journal of Chemical Physics}
  \textbf{\bibinfo{volume}{111}}, \bibinfo{pages}{6197} (\bibinfo{year}{1999}),
  \urlprefix\url{https://doi.org/10.1063/1.479924}.

\bibitem[{\citenamefont{Ayers and Levy}(2001)}]{ayers_levy}
\bibinfo{author}{\bibfnamefont{P.~W.} \bibnamefont{Ayers}} \bibnamefont{and}
  \bibinfo{author}{\bibfnamefont{M.}~\bibnamefont{Levy}}, \bibinfo{journal}{The
  Journal of Chemical Physics} \textbf{\bibinfo{volume}{115}},
  \bibinfo{pages}{4438} (\bibinfo{year}{2001}),
  \urlprefix\url{https://doi.org/10.1063/1.1379333}.

\bibitem[{\citenamefont{Hirata et~al.}(2001)\citenamefont{Hirata, Ivanov,
  Grabowski, Bartlett, Burke, and Talman}}]{Hirata_oep}
\bibinfo{author}{\bibfnamefont{S.}~\bibnamefont{Hirata}},
  \bibinfo{author}{\bibfnamefont{S.}~\bibnamefont{Ivanov}},
  \bibinfo{author}{\bibfnamefont{I.}~\bibnamefont{Grabowski}},
  \bibinfo{author}{\bibfnamefont{R.~J.} \bibnamefont{Bartlett}},
  \bibinfo{author}{\bibfnamefont{K.}~\bibnamefont{Burke}}, \bibnamefont{and}
  \bibinfo{author}{\bibfnamefont{J.~D.} \bibnamefont{Talman}},
  \bibinfo{journal}{The Journal of Chemical Physics}
  \textbf{\bibinfo{volume}{115}}, \bibinfo{pages}{1635} (\bibinfo{year}{2001}),
  \urlprefix\url{https://doi.org/10.1063/1.1381013}.

\bibitem[{\citenamefont{Heßelmann et~al.}(2007)\citenamefont{Heßelmann,
  Götz, Della~Sala, and Görling}}]{Gorling_numerically_stable_oep}
\bibinfo{author}{\bibfnamefont{A.}~\bibnamefont{Heßelmann}},
  \bibinfo{author}{\bibfnamefont{A.~W.} \bibnamefont{Götz}},
  \bibinfo{author}{\bibfnamefont{F.}~\bibnamefont{Della~Sala}},
  \bibnamefont{and} \bibinfo{author}{\bibfnamefont{A.}~\bibnamefont{Görling}},
  \bibinfo{journal}{The Journal of Chemical Physics}
  \textbf{\bibinfo{volume}{127}}, \bibinfo{pages}{054102}
  (\bibinfo{year}{2007}), \urlprefix\url{https://doi.org/10.1063/1.2751159}.

\bibitem[{\citenamefont{Görling et~al.}(2008)\citenamefont{Görling,
  Heßelmann, Jones, and Levy}}]{Gorling_balanced_2}
\bibinfo{author}{\bibfnamefont{A.}~\bibnamefont{Görling}},
  \bibinfo{author}{\bibfnamefont{A.}~\bibnamefont{Heßelmann}},
  \bibinfo{author}{\bibfnamefont{M.}~\bibnamefont{Jones}}, \bibnamefont{and}
  \bibinfo{author}{\bibfnamefont{M.}~\bibnamefont{Levy}}, \bibinfo{journal}{The
  Journal of Chemical Physics} \textbf{\bibinfo{volume}{128}},
  \bibinfo{pages}{104104} (\bibinfo{year}{2008}),
  \urlprefix\url{https://doi.org/10.1063/1.2826366}.

\bibitem[{\citenamefont{Rohr et~al.}(2006)\citenamefont{Rohr, Gritsenko, and
  Baerends}}]{ROHR2006193}
\bibinfo{author}{\bibfnamefont{D.~R.} \bibnamefont{Rohr}},
  \bibinfo{author}{\bibfnamefont{O.~V.} \bibnamefont{Gritsenko}},
  \bibnamefont{and} \bibinfo{author}{\bibfnamefont{E.~J.}
  \bibnamefont{Baerends}}, \bibinfo{journal}{Journal of Molecular Structure:
  THEOCHEM} \textbf{\bibinfo{volume}{762}}, \bibinfo{pages}{193 }
  (\bibinfo{year}{2006}), ISSN \bibinfo{issn}{0166-1280},
  \urlprefix\url{http://www.sciencedirect.com/science/article/pii/S0166128005007785}.

\bibitem[{\citenamefont{Dreizler and Gross}(1990)}]{gross_book}
\bibinfo{author}{\bibfnamefont{R.}~\bibnamefont{Dreizler}} \bibnamefont{and}
  \bibinfo{author}{\bibfnamefont{E.}~\bibnamefont{Gross}},
  \emph{\bibinfo{title}{Density Functional Theory: An Approach to the Quantum
  Many-Body Problem}} (\bibinfo{publisher}{Springer-Verlag},
  \bibinfo{address}{Berlin, Heidelberg}, \bibinfo{year}{1990}), vol.
  \bibinfo{volume}{577} of \emph{\bibinfo{series}{API Conference Proceedings}},
  pp. \bibinfo{pages}{1--20}.

\bibitem[{\citenamefont{Staroverov et~al.}(2006)\citenamefont{Staroverov,
  Scuseria, and Davidson}}]{Stavoverov_OEP}
\bibinfo{author}{\bibfnamefont{V.~N.} \bibnamefont{Staroverov}},
  \bibinfo{author}{\bibfnamefont{G.~E.} \bibnamefont{Scuseria}},
  \bibnamefont{and} \bibinfo{author}{\bibfnamefont{E.~R.}
  \bibnamefont{Davidson}}, \bibinfo{journal}{The Journal of Chemical Physics}
  \textbf{\bibinfo{volume}{124}}, \bibinfo{pages}{141103}
  (\bibinfo{year}{2006}), \urlprefix\url{https://doi.org/10.1063/1.2194546}.

\bibitem[{\citenamefont{Gidopoulos and
  Lathiotakis}(2012{\natexlab{b}})}]{nonanalycity_oep}
\bibinfo{author}{\bibfnamefont{N.~I.} \bibnamefont{Gidopoulos}}
  \bibnamefont{and} \bibinfo{author}{\bibfnamefont{N.~N.}
  \bibnamefont{Lathiotakis}}, \bibinfo{journal}{Phys. Rev. A}
  \textbf{\bibinfo{volume}{85}}, \bibinfo{pages}{052508}
  (\bibinfo{year}{2012}{\natexlab{b}}),
  \urlprefix\url{https://link.aps.org/doi/10.1103/PhysRevA.85.052508}.

\bibitem[{\citenamefont{Friedrich et~al.}(2013)\citenamefont{Friedrich,
  Betzinger, and Bl\"ugel}}]{comment}
\bibinfo{author}{\bibfnamefont{C.}~\bibnamefont{Friedrich}},
  \bibinfo{author}{\bibfnamefont{M.}~\bibnamefont{Betzinger}},
  \bibnamefont{and} \bibinfo{author}{\bibfnamefont{S.}~\bibnamefont{Bl\"ugel}},
  \bibinfo{journal}{Phys. Rev. A} \textbf{\bibinfo{volume}{88}},
  \bibinfo{pages}{046501} (\bibinfo{year}{2013}),
  \urlprefix\url{https://link.aps.org/doi/10.1103/PhysRevA.88.046501}.

\bibitem[{\citenamefont{Gidopoulos and Lathiotakis}(2013)}]{reply}
\bibinfo{author}{\bibfnamefont{N.~I.} \bibnamefont{Gidopoulos}}
  \bibnamefont{and} \bibinfo{author}{\bibfnamefont{N.~N.}
  \bibnamefont{Lathiotakis}}, \bibinfo{journal}{Phys. Rev. A}
  \textbf{\bibinfo{volume}{88}}, \bibinfo{pages}{046502}
  (\bibinfo{year}{2013}),
  \urlprefix\url{https://link.aps.org/doi/10.1103/PhysRevA.88.046502}.

\bibitem[{\citenamefont{Grabo et~al.}(1997)\citenamefont{Grabo, Kreibich, and
  ~}}]{grabo1997}
\bibinfo{author}{\bibfnamefont{T.}~\bibnamefont{Grabo}},
  \bibinfo{author}{\bibfnamefont{T.}~\bibnamefont{Kreibich}}, \bibnamefont{and}
  \bibinfo{author}{\bibfnamefont{E.}~\bibnamefont{~}},
  \bibinfo{journal}{Molecular Engineering} \textbf{\bibinfo{volume}{7}},
  \bibinfo{pages}{27} (\bibinfo{year}{1997}).

\bibitem[{\citenamefont{Engel}(2003)}]{engel2003orbital}
\bibinfo{author}{\bibfnamefont{E.}~\bibnamefont{Engel}}, in
  \emph{\bibinfo{booktitle}{A Primer in Density Functional Theory}}
  (\bibinfo{publisher}{Springer}, \bibinfo{year}{2003}), pp.
  \bibinfo{pages}{56--122}.

\bibitem[{\citenamefont{Uns{\"o}ld}(1927)}]{unsold}
\bibinfo{author}{\bibfnamefont{A.}~\bibnamefont{Uns{\"o}ld}},
  \bibinfo{journal}{Zeitschrift f{\"u}r Physik} \textbf{\bibinfo{volume}{43}},
  \bibinfo{pages}{563} (\bibinfo{year}{1927}).

\bibitem[{\citenamefont{Krieger
  et~al.}(1992{\natexlab{a}})\citenamefont{Krieger, Li, and Iafrate}}]{kli1}
\bibinfo{author}{\bibfnamefont{J.}~\bibnamefont{Krieger}},
  \bibinfo{author}{\bibfnamefont{Y.}~\bibnamefont{Li}}, \bibnamefont{and}
  \bibinfo{author}{\bibfnamefont{G.}~\bibnamefont{Iafrate}},
  \bibinfo{journal}{Physical Review A} \textbf{\bibinfo{volume}{45}},
  \bibinfo{pages}{101} (\bibinfo{year}{1992}{\natexlab{a}}),
  \urlprefix\url{https://link.aps.org/doi/10.1103/PhysRevA.45.101}.

\bibitem[{\citenamefont{Krieger
  et~al.}(1992{\natexlab{b}})\citenamefont{Krieger, Li, and Iafrate}}]{kli2}
\bibinfo{author}{\bibfnamefont{J.}~\bibnamefont{Krieger}},
  \bibinfo{author}{\bibfnamefont{Y.}~\bibnamefont{Li}}, \bibnamefont{and}
  \bibinfo{author}{\bibfnamefont{G.}~\bibnamefont{Iafrate}},
  \bibinfo{journal}{Physical Review A} \textbf{\bibinfo{volume}{46}},
  \bibinfo{pages}{5453} (\bibinfo{year}{1992}{\natexlab{b}}),
  \urlprefix\url{https://link.aps.org/doi/10.1103/PhysRevA.46.5453}.

\bibitem[{\citenamefont{Gritsenko and Baerends}(2001)}]{CEDA}
\bibinfo{author}{\bibfnamefont{O.~V.} \bibnamefont{Gritsenko}}
  \bibnamefont{and} \bibinfo{author}{\bibfnamefont{E.~J.}
  \bibnamefont{Baerends}}, \bibinfo{journal}{Phys. Rev. A}
  \textbf{\bibinfo{volume}{64}}, \bibinfo{pages}{042506}
  (\bibinfo{year}{2001}),
  \urlprefix\url{https://link.aps.org/doi/10.1103/PhysRevA.64.042506}.

\bibitem[{\citenamefont{Della~Sala and G{\"o}rling}(2001)}]{localizedHF}
\bibinfo{author}{\bibfnamefont{F.}~\bibnamefont{Della~Sala}} \bibnamefont{and}
  \bibinfo{author}{\bibfnamefont{A.}~\bibnamefont{G{\"o}rling}},
  \bibinfo{journal}{The Journal of Chemical Physics}
  \textbf{\bibinfo{volume}{115}}, \bibinfo{pages}{5718} (\bibinfo{year}{2001}).

\bibitem[{\citenamefont{Wu and Yang}(2003)}]{Yang_regularized_1}
\bibinfo{author}{\bibfnamefont{Q.}~\bibnamefont{Wu}} \bibnamefont{and}
  \bibinfo{author}{\bibfnamefont{W.}~\bibnamefont{Yang}},
  \bibinfo{journal}{Journal of Theoretical and Computational Chemistry}
  \textbf{\bibinfo{volume}{02}}, \bibinfo{pages}{627} (\bibinfo{year}{2003}),
  \urlprefix\url{https://doi.org/10.1142/S0219633603000690}.

\bibitem[{\citenamefont{Heaton-Burgess
  et~al.}(2007)\citenamefont{Heaton-Burgess, Bulat, and
  Yang}}]{Yang_regularized_2}
\bibinfo{author}{\bibfnamefont{T.}~\bibnamefont{Heaton-Burgess}},
  \bibinfo{author}{\bibfnamefont{F.~A.} \bibnamefont{Bulat}}, \bibnamefont{and}
  \bibinfo{author}{\bibfnamefont{W.}~\bibnamefont{Yang}},
  \bibinfo{journal}{Phys. Rev. Lett.} \textbf{\bibinfo{volume}{98}},
  \bibinfo{pages}{256401} (\bibinfo{year}{2007}),
  \urlprefix\url{https://link.aps.org/doi/10.1103/PhysRevLett.98.256401}.

\bibitem[{\citenamefont{Kollmar and Filatov}(2008)}]{Kollmar_Filatov_OEP}
\bibinfo{author}{\bibfnamefont{C.}~\bibnamefont{Kollmar}} \bibnamefont{and}
  \bibinfo{author}{\bibfnamefont{M.}~\bibnamefont{Filatov}},
  \bibinfo{journal}{The Journal of Chemical Physics}
  \textbf{\bibinfo{volume}{128}}, \bibinfo{pages}{064101}
  (\bibinfo{year}{2008}), \eprint{https://doi.org/10.1063/1.2834214},
  \urlprefix\url{https://doi.org/10.1063/1.2834214}.

\bibitem[{\citenamefont{Pitts et~al.}()\citenamefont{Pitts, Bousiadi,
  Gidopoulos, and Lathiotakis}}]{fscr}
\bibinfo{author}{\bibfnamefont{T.}~\bibnamefont{Pitts}},
  \bibinfo{author}{\bibfnamefont{S.}~\bibnamefont{Bousiadi}},
  \bibinfo{author}{\bibfnamefont{N.~I.} \bibnamefont{Gidopoulos}},
  \bibnamefont{and} \bibinfo{author}{\bibfnamefont{N.~N.}
  \bibnamefont{Lathiotakis}}, \bibinfo{note}{to appear}.

\bibitem[{\citenamefont{Dunning}(1989)}]{Dunning_1}
\bibinfo{author}{\bibfnamefont{T.~H.} \bibnamefont{Dunning}},
  \bibinfo{journal}{The Journal of Chemical Physics}
  \textbf{\bibinfo{volume}{90}}, \bibinfo{pages}{1007} (\bibinfo{year}{1989}),
  \urlprefix\url{https://doi.org/10.1063/1.456153}.

\bibitem[{\citenamefont{Woon and Dunning}(1993)}]{Dunning_2}
\bibinfo{author}{\bibfnamefont{D.~E.} \bibnamefont{Woon}} \bibnamefont{and}
  \bibinfo{author}{\bibfnamefont{T.~H.} \bibnamefont{Dunning}},
  \bibinfo{journal}{The Journal of Chemical Physics}
  \textbf{\bibinfo{volume}{98}}, \bibinfo{pages}{1358} (\bibinfo{year}{1993}),
  \urlprefix\url{https://doi.org/10.1063/1.464303}.

\bibitem[{\citenamefont{Schmidt et~al.}(1993)\citenamefont{Schmidt, Baldridge,
  Boatz, Elbert, Gordon, Jensen, Koseki, Matsunaga, Nguyen, Su
  et~al.}}]{gamess1}
\bibinfo{author}{\bibfnamefont{M.~W.} \bibnamefont{Schmidt}},
  \bibinfo{author}{\bibfnamefont{K.~K.} \bibnamefont{Baldridge}},
  \bibinfo{author}{\bibfnamefont{J.~A.} \bibnamefont{Boatz}},
  \bibinfo{author}{\bibfnamefont{S.~T.} \bibnamefont{Elbert}},
  \bibinfo{author}{\bibfnamefont{M.~S.} \bibnamefont{Gordon}},
  \bibinfo{author}{\bibfnamefont{J.~a.~H.} \bibnamefont{Jensen}},
  \bibinfo{author}{\bibfnamefont{S.}~\bibnamefont{Koseki}},
  \bibinfo{author}{\bibfnamefont{N.}~\bibnamefont{Matsunaga}},
  \bibinfo{author}{\bibfnamefont{K.~A.} \bibnamefont{Nguyen}},
  \bibinfo{author}{\bibfnamefont{S.}~\bibnamefont{Su}}, \bibnamefont{et~al.},
  \bibinfo{journal}{J. Comput. Chem.} \textbf{\bibinfo{volume}{14}},
  \bibinfo{pages}{1347} (\bibinfo{year}{1993}),
  \urlprefix\url{https://onlinelibrary.wiley.com/doi/abs/10.1002/jcc.540141112}.

\bibitem[{\citenamefont{Gordon and Schmidt}(2005)}]{gamess2}
\bibinfo{author}{\bibfnamefont{M.~S.} \bibnamefont{Gordon}} \bibnamefont{and}
  \bibinfo{author}{\bibfnamefont{M.~W.} \bibnamefont{Schmidt}}, in
  \emph{\bibinfo{booktitle}{Theory and Applications of Computational
  Chemistry}}, edited by \bibinfo{editor}{\bibfnamefont{C.~E.}
  \bibnamefont{Dykstra}},
  \bibinfo{editor}{\bibfnamefont{G.}~\bibnamefont{Frenking}},
  \bibinfo{editor}{\bibfnamefont{K.~S.} \bibnamefont{Kim}}, \bibnamefont{and}
  \bibinfo{editor}{\bibfnamefont{G.~E.} \bibnamefont{Scuseria}}
  (\bibinfo{publisher}{Elsevier}, \bibinfo{address}{Amsterdam},
  \bibinfo{year}{2005}), pp. \bibinfo{pages}{1167 -- 1189}, ISBN
  \bibinfo{isbn}{978-0-444-51719-7},
  \urlprefix\url{http://www.sciencedirect.com/science/article/pii/B9780444517197500846}.

\bibitem[{\citenamefont{Pritchard et~al.}(2019)\citenamefont{Pritchard,
  Altarawy, Didier, Gibson, and Windus}}]{bse}
\bibinfo{author}{\bibfnamefont{B.~P.} \bibnamefont{Pritchard}},
  \bibinfo{author}{\bibfnamefont{D.}~\bibnamefont{Altarawy}},
  \bibinfo{author}{\bibfnamefont{B.}~\bibnamefont{Didier}},
  \bibinfo{author}{\bibfnamefont{T.~D.} \bibnamefont{Gibson}},
  \bibnamefont{and} \bibinfo{author}{\bibfnamefont{T.~L.}
  \bibnamefont{Windus}}, \bibinfo{journal}{Journal of Chemical Information and
  Modeling} \textbf{\bibinfo{volume}{59}}, \bibinfo{pages}{4814}
  (\bibinfo{year}{2019}), \bibinfo{note}{pMID: 31600445},
  \urlprefix\url{https://doi.org/10.1021/acs.jcim.9b00725}.

\bibitem[{\citenamefont{Andrade and Aspuru-Guzik}(2011)}]{andrade}
\bibinfo{author}{\bibfnamefont{X.}~\bibnamefont{Andrade}} \bibnamefont{and}
  \bibinfo{author}{\bibfnamefont{A.}~\bibnamefont{Aspuru-Guzik}},
  \bibinfo{journal}{Phys. Rev. Lett.} \textbf{\bibinfo{volume}{107}},
  \bibinfo{pages}{183002} (\bibinfo{year}{2011}),
  \urlprefix\url{https://link.aps.org/doi/10.1103/PhysRevLett.107.183002}.

\bibitem[{\citenamefont{Kraisler and Kronik}(2014)}]{kraisler2014fundamental}
\bibinfo{author}{\bibfnamefont{E.}~\bibnamefont{Kraisler}} \bibnamefont{and}
  \bibinfo{author}{\bibfnamefont{L.}~\bibnamefont{Kronik}},
  \bibinfo{journal}{The Journal of chemical physics}
  \textbf{\bibinfo{volume}{140}}, \bibinfo{pages}{18A540}
  (\bibinfo{year}{2014}).

\bibitem[{\citenamefont{Senjean and Fromager}(2018)}]{PhysRevA.98.022513}
\bibinfo{author}{\bibfnamefont{B.}~\bibnamefont{Senjean}} \bibnamefont{and}
  \bibinfo{author}{\bibfnamefont{E.}~\bibnamefont{Fromager}},
  \bibinfo{journal}{Phys. Rev. A} \textbf{\bibinfo{volume}{98}},
  \bibinfo{pages}{022513} (\bibinfo{year}{2018}),
  \urlprefix\url{https://link.aps.org/doi/10.1103/PhysRevA.98.022513}.

\bibitem[{\citenamefont{Senjean and Fromager}(2020)}]{senjean2020n}
\bibinfo{author}{\bibfnamefont{B.}~\bibnamefont{Senjean}} \bibnamefont{and}
  \bibinfo{author}{\bibfnamefont{E.}~\bibnamefont{Fromager}},
  \bibinfo{journal}{International Journal of Quantum Chemistry} p.
  \bibinfo{pages}{e26190} (\bibinfo{year}{2020}).

\bibitem[{\citenamefont{Guandalini et~al.}(2019)\citenamefont{Guandalini,
  Rozzi, R\"as\"anen, and Pittalis}}]{guandalini}
\bibinfo{author}{\bibfnamefont{A.}~\bibnamefont{Guandalini}},
  \bibinfo{author}{\bibfnamefont{C.~A.} \bibnamefont{Rozzi}},
  \bibinfo{author}{\bibfnamefont{E.}~\bibnamefont{R\"as\"anen}},
  \bibnamefont{and} \bibinfo{author}{\bibfnamefont{S.}~\bibnamefont{Pittalis}},
  \bibinfo{journal}{Phys. Rev. B} \textbf{\bibinfo{volume}{99}},
  \bibinfo{pages}{125140} (\bibinfo{year}{2019}),
  \urlprefix\url{https://link.aps.org/doi/10.1103/PhysRevB.99.125140}.

\bibitem[{\citenamefont{Hodgson et~al.}(2017)\citenamefont{Hodgson, Kraisler,
  Schild, and Gross}}]{Hodgson_ensembles}
\bibinfo{author}{\bibfnamefont{M.~J.~P.} \bibnamefont{Hodgson}},
  \bibinfo{author}{\bibfnamefont{E.}~\bibnamefont{Kraisler}},
  \bibinfo{author}{\bibfnamefont{A.}~\bibnamefont{Schild}}, \bibnamefont{and}
  \bibinfo{author}{\bibfnamefont{E.~K.~U.} \bibnamefont{Gross}},
  \bibinfo{journal}{The Journal of Physical Chemistry Letters}
  \textbf{\bibinfo{volume}{8}}, \bibinfo{pages}{5974} (\bibinfo{year}{2017}),
  \bibinfo{note}{pMID: 29179553},
  \urlprefix\url{https://doi.org/10.1021/acs.jpclett.7b02615}.

\bibitem[{\citenamefont{Kraisler et~al.}(2020)\citenamefont{Kraisler, Hodgson,
  and Gross}}]{kraisler2021kohnsham}
\bibinfo{author}{\bibfnamefont{E.}~\bibnamefont{Kraisler}},
  \bibinfo{author}{\bibfnamefont{M.~J.~P.} \bibnamefont{Hodgson}},
  \bibnamefont{and} \bibinfo{author}{\bibfnamefont{E.~K.~U.}
  \bibnamefont{Gross}}, \emph{\bibinfo{title}{From kohn-sham to many-electron
  energies via step structures in the exchange-correlation potential}}
  (\bibinfo{year}{2020}), \eprint{2008.12029}.

\bibitem[{\citenamefont{Callow et~al.}()\citenamefont{Callow, Pearce, and
  Gidopoulos}}]{ilda}
\bibinfo{author}{\bibfnamefont{T.~J.} \bibnamefont{Callow}},
  \bibinfo{author}{\bibfnamefont{B.~J.} \bibnamefont{Pearce}},
  \bibnamefont{and} \bibinfo{author}{\bibfnamefont{N.~I.}
  \bibnamefont{Gidopoulos}}, \bibinfo{note}{to appear}.

\bibitem[{\citenamefont{Callow et~al.}(2020)\citenamefont{Callow, Lathiotakis,
  and Gidopoulos}}]{invert}
\bibinfo{author}{\bibfnamefont{T.~J.} \bibnamefont{Callow}},
  \bibinfo{author}{\bibfnamefont{N.~N.} \bibnamefont{Lathiotakis}},
  \bibnamefont{and} \bibinfo{author}{\bibfnamefont{N.~I.}
  \bibnamefont{Gidopoulos}}, \bibinfo{journal}{The Journal of Chemical Physics}
  \textbf{\bibinfo{volume}{152}}, \bibinfo{pages}{164114}
  (\bibinfo{year}{2020}), \urlprefix\url{https://doi.org/10.1063/5.0005781}.

\end{thebibliography}

\end{document}